\documentclass[10pt,twocolumn]{research-article-2col}

\templatetype{research-article-2col} 

\usepackage{rotating}  
\usepackage{booktabs}
\usepackage{tablefootnote}
\usepackage{xcolor}
\usepackage{sidecap}

\newcommand{\na}{\textsuperscript{23}Na }    

\newcommand{\hh}{\textsuperscript{1}H }

\newcommand{\hhna}{\textsuperscript{1}H/\textsuperscript{23}Na }

\title{Correlation-weighted \na magnetic resonance fingerprinting in the brain}

\author[1]{Lauren F. O'Donnell}
\author[1,2]{Gonzalo G. Rodriguez}
\author[1]{Gregory Lemberskiy}
\author[1,3,4]{Zidan Yu} 
\author[1]{Olga Dergachyova}
\author[1,5]{Martijn Cloos} 
\author[1,3,*]{Guillaume Madelin} 

\affil[1]{Center for Biomedical Imaging, Department of Radiology, NYU Grossman School of Medicine, New York, NY, USA}
\affil[2]{NMR Signal Enhancement, Max Planck Institute for Multidisciplinary Sciences, G\"ottingen, Germany}
\affil[3]{Vilcek Institute for Graduate Biomedical Studies, NYU Langone Health, New York, NY, USA}
\affil[4]{Department of Medicine, John A. Burns School of Medicine, University of Hawaii, Honolulu, HI, USA}
\affil[5]{Donders Centre for Cognitive Neuroimaging, Donders Institute for Brain, Cognition and Behaviour, Radboud University, Nijmegen, The Netherlands}
\affil[*]{Corresponding author: guillaume.madelin@nyulangone.org}

\begin{abstract}

We developed a new sodium magnetic resonance fingerprinting (\na MRF) method for the simultaneous mapping of $\text{T}_\text{1}$, $\text{T}_\text{2,long}^{*}$, $\text{T}_\text{2,short}^{*}$ and sodium density with built-in $\Delta\text{B}_{1}^{+}$ (radiofrequency transmission inhomogeneities) and $\Delta\text{f}_\text{0}$ corrections (frequency offsets). We based our \na MRF implementation on a 3D FLORET sequence with 23 radiofrequency pulses. To capture the complex spin ${\frac{\text{3}}{\text{2}}}$ dynamics of the \na nucleus, the fingerprint dictionary was simulated using the irreducible spherical tensor operators formalism. The dictionary contained 831,512 entries covering a wide range of  $\text{T}_\text{1}$, $\text{T}_\text{2,long}^{*}$, $\text{T}_\text{2,short}^{*}$, $\Delta\text{B}_\text{1}^{+}$ factor and $\Delta\text{f}_\text{0}$ parameters. Fingerprint matching was performed using the Pearson correlation and the resulting relaxation maps were weighted with a subset of the highest correlation coefficients corresponding to signal matches for each voxel. Our \na MRF method was compared against reference methods in a 7-compartment phantom, and applied in brain in five healthy volunteers at 7 T. In phantoms, \na MRF produced values comparable to those obtained with reference methods. Average sodium relaxation time values in cerebrospinal fluid, gray matter and white matter across five healthy volunteers were in good agreement with values previously reported in the literature. 

\end{abstract}

\dates{\textit{\today}}

\begin{document}

\maketitle

\section{Introduction}

Sodium ions (Na\textsuperscript{+}) plays a critical role in the human body and are invariably linked to the maintenance of ionic homeostasis as well as many physiological and electrochemical processes of metabolism \cite{Madelin2013, Madelin2014}. For this reason, the non-invasive detection of these ions from the nuclear magnetic resonance (NMR) signal of the sodium isotope \na (spin $\frac{\text{3}}{\text{2}}$) has become an important modality for the study of cellular ionic homeostasis and volume regulation, and biochemical status throughout the body \cite{Madelin2013, Madelin2014}. In brain, \na MRI has been used to study neurodegenerative disease \cite{Madelin2013, Madelin2014, Petracca2016, Krahe2022, Mellon2009, Mohamed2021, Haeger2021}, tumor pathology \cite{Madelin2013, Madelin2014, Hashimoto1991, NunesNeto2018} and neurological events, such as stroke \cite{Madelin2014,Boada2012} and traumatic brain injury \cite{Gerhalter2021Na}. \\

The intracellular and extracellular spaces in brain tissues both represent motion-restricted environments that give rise to signal contributions coming from the central and satellite transitions, which are strongly influenced by the quadrupolar interactions of the \na nuclear spin system with its surroundings \cite{Madelin2014}. The central and satellites transitions correspond to the transitions (i.e. single quantum coherences [SQC], which can be detected as NMR signal) between the four energy levels of the spin S = 3/2 of \na placed in a large magnetic field, and which are associated with spin states m\textsubscript{s} = 3/2, 1/2, -1/2 and  -3/2, from lowest to highest energy levels, respectively \cite{Madelin2014, Rooney1991}. Transitions (3/2)$\leftrightarrow$(1/2) and (-1/2)$\leftrightarrow$(-3/2) are called “satellite” and transition (1/2)$\leftrightarrow$(-1/2) is called “central”. These transitions occur with different relaxation times, the shorter ones corresponding to the satellite transitions in solids, or in biological tissues with motional restriction for the sodium ions (such as intracellular and extracellular spaces). Since a \na NMR or MRI experiment detect all SQCs related to these transitions, the overall \na signal acquired in most tissues will exhibit a biexponential transverse relaxation decay, \textit{i.e.} a long component ($\text{T}_\text{2,long}^{*}$) and a short component ($\text{T}_\text{2,short}^{*}$), originating in both the intracellular and extracellular spaces. As a consequence, both gray matter (GM) and white matter (WM) in brain will also exhibit an overall \na biexponential transverse relaxation, as a weighted average of the relaxation processes from the intracellular and extracellular spaces. Similarly, \na longitudinal relaxation follows the same biexponential pattern from the intracellular and extracellular spaces. However, in soft biological tissues, both the short and long components are often very close to each other and longitudinal relaxation is usually measured as a monoexponential process ($\text{T}_\text{1,short}\sim\text{T}_\text{1,long}\sim\text{T}_\text{1}$) in GM and WM\cite{Madelin2014}. In fluids such as cerebrospinal fluid (CSF), the quadrupolar interaction averages to zero as a result of rapid motion allowing the signal dynamics to be often modeled as a monoexponential relaxation for both transverse and longitudinal magnetization components. Collectively, the spin ${\frac{\text{3}}{\text{2}}}$ nature of the \na nucleus and the low concentration of Na\textsuperscript{+} ions in brain tissue (on the order of 40-50 mM on average) combined with the inhomogeneous structure of the brain, makes it difficult to simultaneously quantify \na relaxation times and density \cite{Madelin2013,Madelin2014,Lee2009}. 

Proton magnetic resonance fingerprinting (\hh MRF) has become a popular technique for the simultaneous quantification of physical properties within a system \cite{Ma2013, Cloos2016, Panda2017, Mehta2019, Hsieh2020}. Recently, sodium MRF (\na MRF) studies in the brain have demonstrated promising initial results. Kratzer et al. \cite{Kratzer2020, Kratzer2021} implemented a version of \na MRF capable of quantifying relaxation parameters in CSF and brain tissue (combined GM and WM) that utilized a 3D radial sequence with variable repetition times (TR), echo times (TE) and flip angles (FA). Our group previously introduced a multipulse approach to multicompartmental Na\textsuperscript{+} concentration quantification \cite{Gilles2017}, which we now expanded for quantifying \na relaxation in the brain.  

In this work, we present a \na MRF technique that is sensitive enough to quantify differences in average relaxation times over whole GM, WM, and CSF. Our method simultaneously maps \na $\text{T}_\text{1}$, $\text{T}_\text{2,long}^{*}$, $\text{T}_\text{2,short}^{*}$, sodium density (SD), and experimental imperfections arising from radiofrequency (RF) transmission inhomogeneities ($\Delta\text{B}_{1}^{+}$ factor) and frequency offsets from B\textsubscript{0} inhomogeneities ($\Delta\text{f}_{0}$). The \na MRF pulse train with variable FAs and phase angles (PA) was designed by incorporating the irreducible spherical tensor operator (ISTO)\cite{Madelin2014, Lee2009, Gilles2017} formalism into a genetic algorithm (GA) that minimizes signal correlation between GM and WM, assuming average relaxation times from the literature for these two tissues \cite{Nagel2011, Niesporek2017, Blunck2018} during this optimization phase. A 3D Fermat looped orthogonally encoded trajectory (FLORET) \cite{Pipe2011} was used to fully sample k-space with constant TE. The proposed \na MRF sequence can acquire data over the full brain with 5-mm isotropic resolution in about 30 min at 7 T. We tested our \na MRF approach in a 7-compartment phantom and in 5 healthy volunteers. 


\section{Material and Methods}

\subsection*{\na spin dynamics simulation}

The dynamics of the \na spin $I = \frac{\text{3}}{\text{2}}$ were modeled using the ISTO framework under the conventions described in Madelin et al. \cite{Madelin2014}, Lee et al. \cite{Lee2009}, and Gilles et al. \cite{Gilles2017}. Within this formalism, the evolution of the \na spin system is described by the Liouville-von Neumann (master) equation (with convention $\hbar\equiv$ 1): 
\vspace{-4pt}
\begin{equation}
     \frac{d}{dt}\rho(t) = -i\big[ H, \rho(t) \big] - \hat{\Gamma}\{\rho(t) - \rho^{eq}\},
    \label{MasterEqn}  
    \vspace{4pt}
\end{equation} 
where $\rho^{eq}$ is the density operator of the spin system at thermal equilibrium, $H$ is the total spin Hamiltonian and $\hat{\Gamma}$ is the Redfield relaxation superoperator. The total Hamiltonian $H$ is the sum of the main Hamiltonians acting on the density operator, such as the Zeeman Hamiltonian $H_Z$ (interaction of the spins with the constant $B_0$ field), the residual quadrupolar interaction Hamiltonian $H_Q$ (interaction of the quadrupole moment of the nuclei with the residual average electric field gradient in anisotropic environments), and the RF field Hamiltonian $H_1(t)$ (interaction of the spins with time-varying transmit RF field $B_1^+(t)$) \cite{Madelin2014}. According to recent studies \cite{Stobbe2016, Gast2023}, this residual $H_Q$ can actually have an effect on $B_1^+$ efficiency and therefore on the results from quantitative sodium imaging. Since we will include $\Delta B_1^+$ inhomogeneities in our spin dynamics simulation to generate the fingerprint dictionary, we can assume that this anisotropic residual $H_Q$ is negligible in soft quasi-isotropic brain tissues in our model \cite{Madelin2014}. Since we are also operating in the rotating frame for the spin dynamics simulation, the main Hamiltonians acting on the spin system are $H_Z$ related to $B_0$ inhomogeneities only, and $H_1(t)$. 

The Redfield relaxation superoperator acting on the density operator of the spin system is described by:
\vspace{-8pt}
\begin{equation}
\begin{split}
    \hat{\Gamma}\{\rho-\rho^{eq}\} = A\sum^{2}_{m=-2}(-1)^m[T_{2,m},[T_{2,-m},\rho-\rho^{eq}]] \\ \times \big(J(m\omega)-iK(m\omega)\big),
\label{RelaxSup}
\end{split}
\vspace{-4pt}
\end{equation}
where $A$ is a constant that depends on the convention used to describe spectral densities, $T_{2,m}$ is the ISTO of rank 2 and order \textit{m}, $J(m\omega)$ is the spectral density function, and $K(m\omega)$ is the imaginary term associated with dynamic frequency shift, which in our case can be omitted due to its negligible observable effect in soft tissues \cite{Madelin2014}. The matrix formulation of $\hat{\Gamma}$ and relaxation rates $R_i = \frac{1}{T_i}$ with $i$ = \textit{(1,short), (1,long), (2,short), (2,long)}, are given by \cite{Lee2009, Madelin2014}:
\begin{align}
    R_{1,short} & =  6J(0) \label{R1s} \\ 
    R_{1,long}  & =  6J(\omega) \label{R1l} \\ 
    R_{2,short} & =  3J(0) + 3J(\omega) \label{R2s} \\
    R_{2,long}  & =  3J(\omega) + 3J(2\omega).\label{R2l}
\end{align}
When simulating the spin dynamics for \na MRF pulse train optimization or for generating the final fingerprint dictionary, the algorithm takes the relaxation times as input, then calculates the spectral density functions according to Equations \ref{R1s}-\ref{R2l} and uses the results to construct the Redfield relaxation superoperator in Equation \ref{RelaxSup}, which is then added to the Liouville equation. After each time step of the simulation (100 $\mu$s), the simulated \na MR signal, which corresponds to the MR-observable transverse magnetization, is calculated as the average rank-1 single quantum coherence $T_{1,-1} = \frac{1}{\sqrt{2}}I_{-} = \frac{1}{\sqrt{2}}(I_{x}-iI_{y})$ using the standard formula $\langle T_{1,-1} \rangle = \text{Tr}(\rho T_{1,-1})$, where $\text{Tr}(A)$ is the trace of matrix $A$.

\begin{figure*}[t!]
    \centering
    \includegraphics[width=\textwidth]{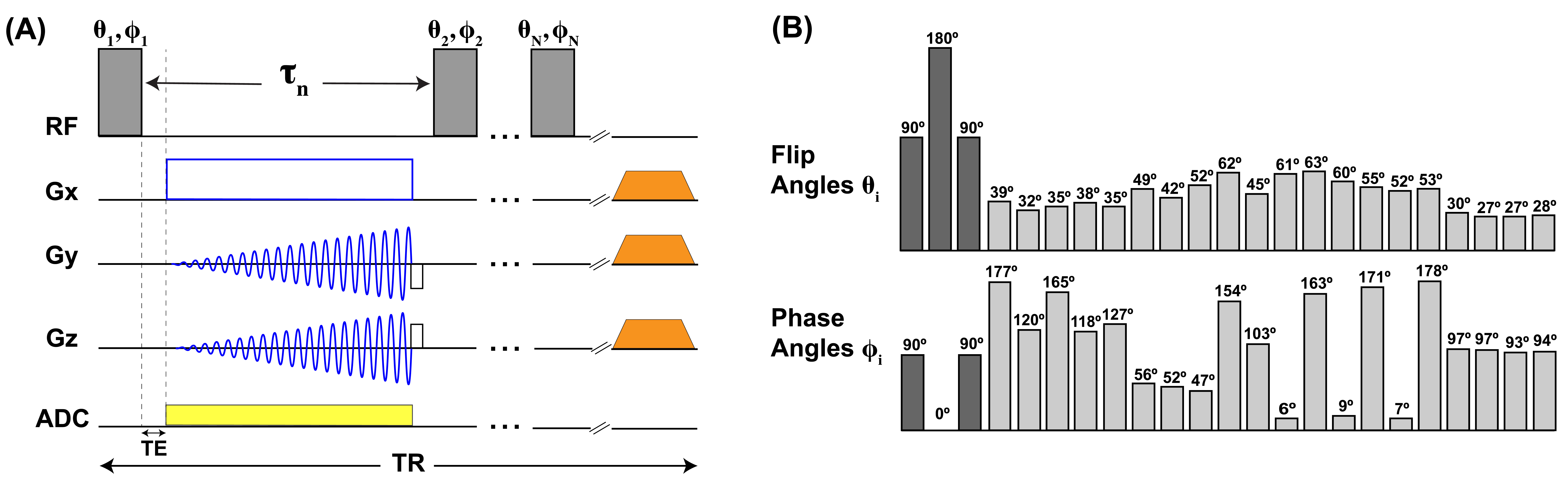} 
    \caption{\textbf{Pulse sequence diagram for 3D \na MRF.} The overall pulse scheme is shown in (\textbf{A}). The diagram in (\textbf{B}) represents the variable flip angle (FA) and phase angle (PA) MRF train. There were a total of $N$ = 23 non-selective rectangular RF pulses of duration $\tau_{RF} = 0.8$ ms. Each pulse was followed by a time period $\tau_{i}$. On the RF channel in (\textbf{A}), the gray rectangles correspond to a single FA = $\theta_{i}$ and PA = $\phi_{i}$ combination in train (\textbf{B}). In (\textbf{B}), the dark gray rectangles represent a 3-pulse inversion composite block \cite{Levitt1986} used to increase $\text{T}_\text{1}$ sensitivity of the sequence and improve RF homogeneity for the magnetization inversion. Within the composite block, $\tau_{1}$ = $\tau_{2}$ = 7.5 ms. The light gray rectangles indicate the 20-pulse variable FA and PA train. All FA = $\theta_{i}$ and PA = $\phi_{i}$ are listed at the top of each rectangle and $\tau_{i}$ = 15 ms for $i \geq 3$. During each time delay $\tau_{i}$, a time period TE = 0.2 ms was followed by the ADC event, indicated by a yellow block on the ADC channel. A 3D spiral encoding scheme using the FLORET trajectory \cite{Gilles2017,Pipe2011,Robinson2017} (3 hubs at 45$^{\circ}$, 100 interleaves/hub), indicated in blue, was played over the gradient channels. Immediately after the ADC, a rewind gradient was also played out and then the next RF pulse in the MRF train was initiated. After $N$ pulses and $N$ delays of duration $\tau_{i}$ have played out, a 5-ms spoiling gradient at 70\% maximum gradient strength, indicated by an orange trapezoid, was applied in all directions. The entirety of this scheme represented one TR of the sequence.} 
    \label{pulseDiagram}
\end{figure*}

\subsection*{Pulse sequence design for 3D \na MRF}

Figure \ref{pulseDiagram}(A) presents the \na MRF sequence. System excitation was driven by a series of $N$ non-selective rectangular RF pulses with FA $\theta_i$ and PA $\phi_i$ followed by a delay time $\tau_{i}$ ($i = 1,2,...,N$), forming the MRF pulse train shown in Figure \ref{pulseDiagram}(B). An initial magnetization inversion using a 90$^\circ$--180$^\circ$--90$^\circ$ (dark gray) composite 3-pulse block \cite{Levitt1986} was used to increase $\text{T}_\text{1}$ sensitivity and improve RF homogeneity for the inversion before initiating the 20-pulse variable FA/PA train (light gray). 

The RF pulse durations were fixed at $\tau_{RF}$ = 0.8 ms, and interpulse delay periods were fixed at $\tau_{i}$ = 7.5 ms within the composite block ($i = 1,2$) and $\tau_{i}$ = 15 ms for the next pulses ($i \geq 3$).  The interpulse delays were set according to our prior work on simultaneous \hhna MRI \cite{Yu2021, Rodriguez2022} and in anticipation of integrating this method into simultaneous \hhna MRF. 

The FLORET trajectory \cite{Gilles2017, Pipe2011, Robinson2017} was used to readout the signal followed by a rewinder to balance the gradient moment. After $N$ pulses and $N\tau_{i}$ delays, a spoiling gradient (duration = 5 ms, 70\% of maximum gradient strength) was applied simultaneously in all directions to ensure complete dephasing of residual transverse magnetization prior to beginning the next TR. A delay of $\sim$300 ms was inserted between RF pulse trains to allow recovery of the longitudinal magnetization and reduce specific absorption rate (SAR). A constant TE = 0.2 ms was used throughout the sequence. 

The variable 20-pulse FA/PA train was optimized using a genetic algorithm (GA) implemented in the Global Optimization Toolbox in MATLAB R2020b (The MathWorks Inc., Natick, Massachusetts, USA) on an Apple MacBook Pro (2019) laptop with a 2.4 GHz 8-Core Intel Core i9 processor. The GA operates by evolving a population of candidate solutions through the processes of selection, recombination and mutation or crossover. First, a random population is initialized and each individual is evaluated using the objective function. These individuals are scored and assigned a fitness value which is then converted into an expectation value. Next, parents for the following population are determined based on their expectation values. At each step, the GA generates a new population of offsprings from the pool of current parents. The GA terminates when the user-defined end criteria are met. 

For the pulse train optimization, the GA minimized an objective function estimating the Pearson correlation (PC) coefficient \cite{Saidi2019} between \na MR signals arising from GM and WM. The ISTO simulation was performed first using input tissue relaxation times based on average values reported in the literature \cite{Madelin2013, Madelin2014, Coste2019}: for GM, $\text{T}_\text{1}$ = 30.2 ms, $\text{T}_\text{2,long}^{*}$ = 26.4 ms, $\text{T}_\text{2,short}^{*}$ = 4.0 ms; for WM, $\text{T}_\text{1}$ = 29.2 ms, $\text{T}_\text{2,long}^{*}$ = 22.1 ms and $\text{T}_\text{2,short}^{*}$ = 3.9 ms. The system was then propagated under a 23-pulse train with user defined starting values for the $N$-pulse train (where $N_{pulse}$ = {4, ..., 23} and $i$ = 1, ..., $N_{pulse}$) were $\theta_{i}$ = 35$^{\circ}$ and $\phi_{i}$ = 0$^{\circ}$ with period $\tau_{i}$ = 15 ms. The composite block, corresponding to $i$ = 1, 2, 3 with $\tau_{1} = \tau_{2}$ = 7.5 ms, was also included in the simulation as a non-variable parameter. Limitations were imposed on FA (0$^{\circ}\leq\theta_{i}\leq$70$^{\circ}$) and PA (0$^{\circ}\leq\phi_{i}\leq$180$^{\circ}$) in consideration of SAR limits. Inputs for the objective function were the real number vectors of the simulated signals of GM and WM. The objective function then determined the PC coefficient between these 2 signals. The GA minimized the relationship between GM and WM by selecting signals for which the PC coefficient was a minimum. Optimizing the system for signals that are less related, according to the PC, allowed for better separation between GM and WM. PC was used in the objective function because of its robustness for determining the strength of real valued quantitative variables and because it can be applied to large datasets \cite{chen2002}. The GA optimized the system for 100 generations in 8 h. The algorithm was applied a total of 3 times, first using the initial input values for the variable 20-pulse part of the pulse train, followed by two iterations in which the solution of the previous computation was taken as input for the next one. This way, the optimized pulse train could be inspected after every 100\textsuperscript{th} generation cycle. 

\subsection*{Fingerprint dictionary simulation}

Simulation of the fingerprint dictionary was performed in MATLAB R2020b on a Cray CS500-1211 cluster with Intel Xeon Gold 6148 high memory CPUs at the NYU Langone High Performance Computing Core facility (New York City, NY, USA). The simulation code is freely available in Matlab File Exchange (see Data Availability section for the link). Signals were simulated starting from thermal equilibrium and propagated under the optimized 23-pulse \na MRF train (Figure \ref{pulseDiagram}(B)). Parameter ranges ([begin:step:end]) to build the dictionary were $\text{T}_\text{1,long}$ = [20:2:74] ms, $\text{T}_\text{1,short}$ = [20:2:74] ms, $\text{T}_\text{2,long}^{*}$ = [10:2:66] ms, $\text{T}_\text{2,short}^{*}$ = [0.5:0.5:2.0, 2:2:66] ms, $\Delta\text{B}_\text{1}^{+}$ factor = [0.7:0.1:1.3] (applied as a multiplying factor to the RF amplitude FA) and $\Delta\text{f}_\text{0}$ = [-60:10:60] Hz. For $\text{T}_\text{1}$ quantification, we originally assumed that T\textsubscript{1,long} = T\textsubscript{1,short}. To insure that spectral densities where $J(0) \neq J(\omega)$ were included in the dictionary, a constant ($\pm\Delta\text{T}_\text{1}$) that added 1 ms to every entry for $\text{T}_\text{1,long}$ and subtracted 1 ms to every entry for $\text{T}_\text{1,short}$ was included in the simulations. Parameter combinations where $\text{T}_\text{2,long}^{*}>\text{T}_\text{1}$ and $\text{T}_\text{2,short}^{*}>\text{T}_\text{2,long}^{*}$ were omitted from the computation. In generating the dictionary, we did not make additional assumptions regarding he ratios between relaxation processes for the central and satellite transitions of the \na spins $\frac{\text{3}}{\text{2}}$ (i.e. the long and short relaxation components). Neither the dictionary nor the ISTO simulation assume an ideal ratio of 0.6:0.4 for $\text{T}_\text{2,short}^{*}$:$\text{T}_\text{2,long}^{*}$ due to quadrupolar relaxation processes. We also note here that we are defining the biexponential $\text{T}_\text{2}^{*}$ parameters compiled in the dictionary as the effective $\text{T}_\text{2}$\cite{Levitt2008}. While $\text{T}_\text{2}$ represents decay of the transverse magnetization of the tissue alone, $\text{T}_\text{2}^{*}$ includes field inhomogeniety effects\cite{Levitt2008,Tang2014}. In this case, we included $\Delta\text{B}_\text{1}^{+}$ factor and $\Delta\text{f}_\text{0}$ explicitly in the dictionary to account for the effects of dephasing contributions from field inhomogeneity. Due to the memory expense incurred by the simulation, the dictionary was generated in two parts and concatenated. In total, 831,512 entries were generated. The total simulation time was 6.4 days.

\subsection*{Experiments}

All experiments were performed at 7 T (MAGNETOM, Siemens, Erlangen, Germany) using a 16-channel transmit/receive \hhna RF brain coil constructed in-house (8 \hh channels, 8 \na channels) \cite{Wang2021}. At the beginning of each scanning session a localizer scan was acquired followed by standard automatic $\text{B}_\text{0}$ shimming from the scanner using the \hh channel.
 
\subsubsection*{Phantom} 

Our test phantom was constructed using a 2.3-L cylinder (outer diameter OD = 20 cm, length = 35 cm) filled with a solution of 70 mM NaCl and which contained seven 50-mL polypropylene cylinders (OD = 30 mm, length = 115 mm) arranged as 6 outer samples each containing a different concentration of NaCl and agar circling the 7\textsuperscript{th} sample placed in the middle and containing a solution of 140 mM NaCl. NaCl concentrations, and thus sodium density ground truth values in the phantoms, were estimated to have a 5\% uncertainty due to sample preparation, originating from uncertainty of the scale used to weigh the water, NaCl and Agar before heating and mixing.

\begin{figure*}[t!]
    \centering
    \includegraphics[width=1\textwidth]{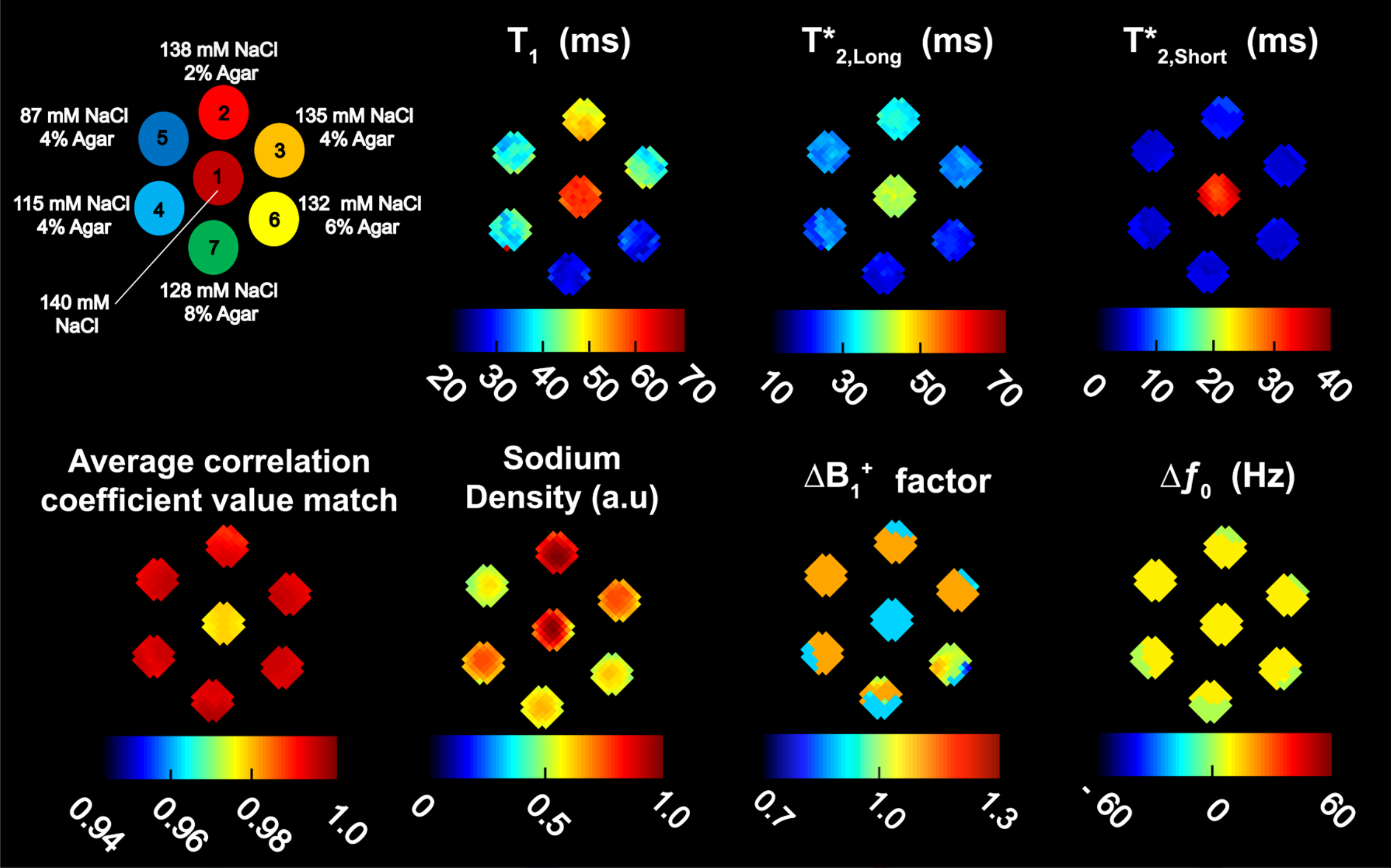}
    \caption{\textbf{\na MRF maps of the 7-compartment phantom.} Fingerprint matching was performed over an average of 2 center axial slices with 20 correlation coefficients included in the weighted average. A diagram of the 7-compartment phantom is shown in the top left corner. Maps for $\text{T}_\text{1}$, $\text{T}_\text{2,long}^{*}$ and $\text{T}_\text{2,short}^{*}$ are shown on the top row. A map showing the average correlation coefficient values matched for 20 correlations is shown at bottom left, followed by maps for normalized SD in arbitrary units (a.u), $\Delta\text{B}_\text{1}^{+}$ factor (unitless) and $\Delta\text{f}_\text{0}$ in Hz.}
    \vspace{-4pt}
    \label{PhantomFig}
\end{figure*}

A diagram of the phantom with sodium/agar concentrations in the axial orientation is shown in Figure \ref{PhantomFig}. The regions-of-interest (ROI) in the 7 samples were generated from a 3D mask of equal diameter to each sample such that all ROI volumes would be the same. 

For \na MRF, one scan consisting of 16 averages with TR = 511 ms, isotropic resolution = 5 mm and isotropic FOV = 320 mm, was acquired with the FLORET parameters: 3 hubs/45$^{\circ}$ with 100 interleaves/hub, TE = 0.2 ms, readout duration = 5 ms within the composite at the beginning of the pulse train ($\tau_i$ = 7.5 ms for $i=1,2$) and 10.230 ms after the next pulses ($\tau_i$ = 15 ms for $i\geq3$), total scan time = 40:52 min. 

We measured the reference $\text{T}_\text{1}$ using a saturation recovery (SR) reference experiment that consisted of a series of eight scans. Each scan consisted of a FLORET sequence (3 hubs/45$^{\circ}$ with 100 interleaves per hub) with different TRs = [60, 100, 140, 180, 260, 300, 360, 420] ms and fixed TE = 0.1 ms, 4 averages, isotropic resolution = 5 mm, isotropic FOV = 320 mm, rectangular RF excitation pulses with FA = 90$^\circ$ and $\tau_{RF}$ = 0.8 ms. The total scan time for the SR experiments running sequentially was 1:49 h.

Similarly, we measured the reference mono- and biexponential $\text{T}_\text{2}^*$ using a multi-TE experiment that consisted of a series of 13 scans. Each scan consisted of a FLORET sequence (3 hubs/45$^{\circ}$ with 100 interleaves per hub) with TEs = [0.1, 0.5, 1.0, 1.5, 2, 5, 7.5, 10, 15, 25, 35, 50, 65] ms and fixed TR = 150 ms, 4 averages, isotropic resolution = 5 mm, isotropic FOV = 320 mm, rectangular RF excitation pulses with FA = 90$^\circ$ and $\tau_{RF}$ = 0.8 ms. The total scan time for the multi-TE experiments running sequentially was 1:57 h.

\subsubsection*{Brain} 

Five healthy volunteers (1 female, 4 males, mean age 36 $\pm$ 8.5 years) were recruited under a protocol approved by the New York University Grossman School of Medicine institutional review board. All parts of the study were performed in accordance with the relevant guidelines and regulations set forth by the Human Research Protections Program. Informed consent was obtained before each scanning session. For \na MRF, we acquired 4 separate scans consisting of 2 averages per scan. These 4 scans were acquired consecutively within 29.8 $\pm$ 1.3 min. We chose to divide the \na MRF scanning portion of the session into separate 2-average scans in order to communicate with the volunteers during the acquisition. The average TR over all volunteer scans was TR = 704 $\pm$ 4 ms, which varied between subjects due to head size and coil loading. The minimum TR was calculated by the scanner to keep SAR within the maximum limit of 100\%. Setting TR $>$700 ms allowed for full $\text{T}_\text{1}$ recovery for about 380 ms after the last RF pulse, which was more than 5 times the longest expected \na $\text{T}_\text{1}$ in brain from CSF ($\text{T}_\text{1}\sim$ 50-60 ms \cite{Madelin2014, Kratzer2021}). All brain scans were acquired with isotropic resolution = 5 mm, isotropic FOV = 320 mm, and FLORET parameters: 3 hubs/45$^{\circ}$ with 100 interleaves/hub, TE = 0.2 ms, readout duration = 5 ms within the composite at the beginning of the pulse train (where $\tau_i$ = 7.5 ms for $i=1,2$) and 10.230 ms after the next pulses (where $\tau_i$ = 15 ms for $i\geq3$).

For reference, a \hh MPRAGE was acquired with 1 average, TR = 2300 ms, TE = 2.84 ms, FOV = 256$\times$216 mm\textsuperscript{2}, slice thickness = 1 mm, 176 slices/slab and GRAPPA acceleration factor 2, for a total acquisition time of 4:32 min. The average total session time for calibration, shimming and scanning across all volunteers was 55 $\pm$ 3 min.  

\subsection*{Data processing} 

All images were reconstructed offline in MATLAB. For each channel, raw k-space data acquired during the \na MRF sequence was filtered with a Hamming kernel then reconstructed using gridding \cite{Pipe1999, Zwart2012} to produce a set of 23 complex images with a final nominal resolution 2.5$\times$2.5$\times$2.5 mm\textsuperscript{3} (128$\times$128$\times$128 matrix). The images from the 8 channels were combined using coil sensitivity profiles as described by Bydder et al. \cite{Bydder2002}. The average time for image reconstruction in the brain across all five volunteers was 42 s, and 38 s in the phantom. For the brain images, an additional denoising step was performed on the complex images using the Marchenko-Pastur method \cite{Veraart2016, Lemberskiy2019, Lemberskiy2020}. This added another 15 s to the brain image reconstruction time. Supplementary Figure S1 shows 23 axial images for a center slice in the phantom. Supplementary Figures S2 and S3 show 23 reconstructed axial images for a center slice in the brain of volunteer 5.

Phantom images from FLORET for $\text{T}_\text{1}$ and $\text{T}_\text{2}^*$ reference experiments were reconstructed in the same way as the \na MRF data. Curve fitting was applied voxelwise over the axial plane in the central slice of the phantom using the Levenberg-Marquardt algorithm applied using \textit{lsqcurvefit} in MATLAB. A monoexponential kernel was assumed for $\text{T}_\text{1}$ \cite{Gast2023} according to Equation \ref{monoEx}: 
\begin{equation}
    S\,(\text{TR},\,\text{T}_{1}) = B\sqrt{\left(1 - \,e^{-\frac{\text{TR}}{\text{T}_{1}}}\,\right)^{2}+\,\mathcal{N}^{2}}, 
    \label{monoEx}      
\end{equation}
where coefficient $B$ and noise floor $\mathcal{N}$ were variable over the fit optimization. $\text{T}_\text{1}$ values were restricted to a lower bound of 20 ms and upper bound of 80 ms, and TR was the repetition time from the FLORET acquisitions. Time required for this process was 33 s.

\begin{figure*}[t!]
    \centering
    \includegraphics[width=\textwidth]{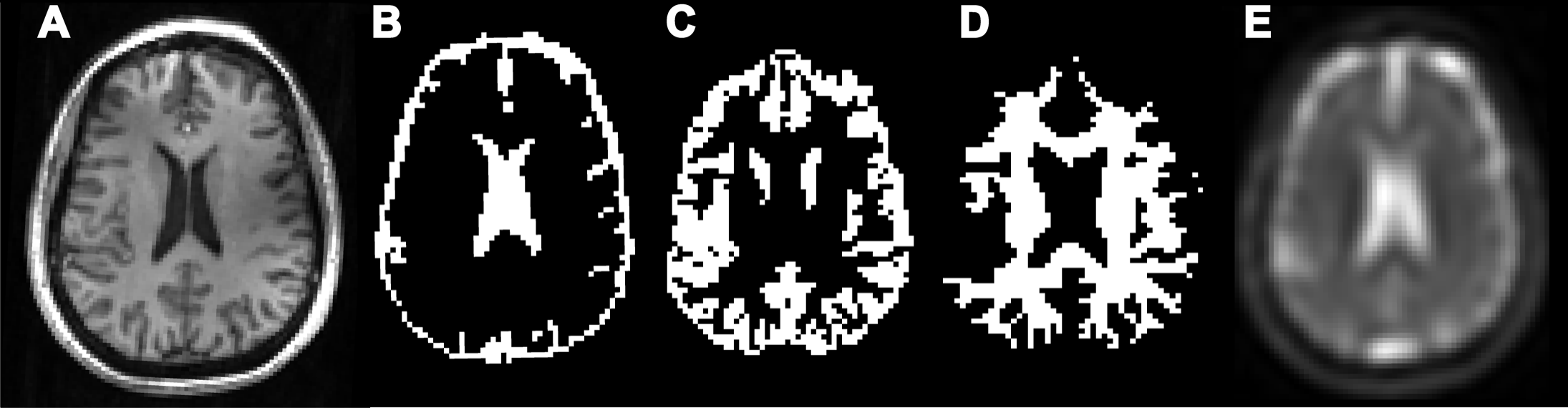}
    \caption{\textbf{Images of a center slice of the brain in the axial position from volunteer 5.} (\textbf{A}) \hh MPRAGE.  (\textbf{B}) Binary mask for CSF. (\textbf{C}) Binary mask for GM. (\textbf{D}) Binary mask for WM. (\textbf{E}) \na image acquired after first RF pulse of the MRF pulse train.}
    \label{masks}
\end{figure*}

For $\text{T}_\text{2}^{*}$, we applied the biexponential kernel shown in Equation \ref{biEx} as descibed by Ridley et. al \cite{Ridley2018,Gast2023}. 
\begin{equation}
\begin{split}
    S(\text{TE},\,\text{T}^{*}_{2}) = \\ 
    \sqrt{A^{2} \left(f \cdot \,e^{-\frac{\text{TE}}{\text{T}^{*}_\text{2,short}}}\,+\, (1-f)\,e^{-\frac{\text{TE}}{\text{T}^{*}_\text{2,long}}}\right)\,+\,n^{2} },
    \label{biEx} 
\end{split}
\end{equation}
The amplitude scaling factor $A$ was initialized as 1 and allowed to vary during the optimization with the lower bound set at 0 and the upper bound set at infinity. The Rician noise parameter $n$ was initialized as 0.1 based on the average noise measured in a background region outside of the phantom on the magnitude image, as described in Qian et al. \cite{Qian2021}. In this case, the upper bound was set at 1 with a lower bound of 0.01. The sodium signal fraction $f$ was initialized at 0.6 and allowed to vary between a lower bound of 0.4 and an upper bound of 0.8. The TE variable corresponded to TEs from the FLORET acquisitions. $\text{T}_\text{2,short}^{*}$ was bound between 0.5 ms and 60 ms and $\text{T}_\text{2,long}^{*}$ was bound between 10 ms and 80 ms.

In both fits, the upper and lower bounds for relaxation times were chosen to coincide with those parameter ranges simulated in the \na MRF dictionary. Finally, the average relaxation values from the 7 individual phantom samples were calculated from the resultant 2D relaxation maps after masking. The time required for fitting was 27 s. Collectively, this process was referred to as the reference method (RM).

For each volunteer, images from the \hh MPRAGE DICOM datasets were co-registered to the \na MRF data using SPM12 (UCL, London, UK) \cite{SPM12, Rodriguez2022}. The \na MRF image used for co-registration was the one acquired just after the first pulse in the \na MRF train. Tissue probability maps were generated from the normalized and co-registered MPRAGE images and segmented into CSF, GM and WM tissue classes using SPM12 \cite{SPM12, Rodriguez2022}. The segmentated regions for CSF, GM and WM were then binarized with a 90\% threshold to generate tissue ROI masks in MATLAB. These images are presented in Figure \ref{masks}. The high threshold was chosen to reduce the likelihood of contamination between different tissues.

\subsection*{Fingerprint dictionary matching} 

The fingerprint dictionary (size = 831,512 entries) was matched voxelwise to the complex \na image data using Pearson correlation. In the phantom, matching was performed on the average of two center slices in the axial plane. Matching required 10:34 min per slice followed by reconstruction of the correlation-weighted maps which added another 10 s. In brain, matching was performed over one center slice in each of the axial, coronal and sagittal positions for all five volunteers. Additionally, a slab of 20 axial slices were matched for volunteer 5. Matching required an average of 9:42 min per slice plus 10 s for reconstruction of the correlation-weighted maps. All these processes were also performed on an Apple MacBook Pro (2019, 2.4 GHz 8-Core Intel Core i9).

Because of the dictionary size, and due to the low SNR of the sodium images, it was possible that more than one match could generate a high correlation for a single voxel. To account for this, we included matches for a subset of the top correlations for each voxel $v$, and generated the final maps by calculating the correlation-weighted parameter $X_{v}$ from the dictionary of values $X_{v,i}$ corresponding to the matching correlation coefficient $w_{v,i}$ according to:  
\begin{equation}
    X_{v} = \frac{\sum^{k}_{i=1} w_{v,i}X_{v,i}}{\sum^{k}_{i=1} w_{v,i}},
    \label{whAve} 
\end{equation}
where $k$ was the maximum number of correlation coefficients used for weighting and $X_{v,i}$ and $X_{v}$ were the unweighted and weighted parameters $\text{T}_\text{1}$, $\text{T}_\text{2,short}^{*}$, $\text{T}_\text{2,long}^{*}$, $\Delta\text{B}_{1}^{+}$ factor or $\Delta\text{f}_{0}$, respectively.

SD in voxel $v$ was calculated as the mean ratio between the mean absolute value of the measured fingerprint in the voxel and the mean absolute value of corresponding fingerprint from the dictionary with correlation coefficient $w_{v,i}$. Once SD was calculated for each voxel $v$ and each correlation coefficient $w_{v,i}$, the resulting SD maps for each $w_{v,i}$, which are now in arbitrary units depending on the data reconstruction algorithm and hardware characterisitics (RF chain, preamplifiers, ADC) were then normalized by dividing them by the highest voxel value in the whole 3D SD dataset. Weighting was then done according to Equation \ref{whAve}, where $X_{v,i}$ was the unweighted SD and $X_{v}$ was the weighted SD. Weighting the SD maps by the correlation coefficients reduces the effect of noise in both the dictionary matching and the normalization processes used to generate the SD maps.

\begin{table*}[t!]
    \centering
    \caption{\textbf{\na relaxation times measured in the 7-compartment phantom.} Measurements are shown as mean value $\pm$ standard deviation, from our \na MRF method and from the average of two repetitions of the reference method. \label{phantomTable}}
    \resizebox{\textwidth}{!}{
    \scriptsize
    \begin{tabular}{lllllllllll}
    \hline 
    \multicolumn{1}{l}{} & \multicolumn{2}{l}{\textbf{Concentration}} & & \multicolumn{3}{l}{\textbf{$^{23}$Na MRF}} & & \multicolumn{3}{l}{\textbf{Reference Method}} \\
    \cline{2-3} \cline{5-7} \cline{9-11}
    \textbf{ROI} & \textbf{Agar (\%)} & \textbf{NaCl (mM)} & & \textbf{$\text{T}_\text{1}$ (ms)} & \textbf{$\text{T}_\text{2,long}^{*}$ (ms)} & \textbf{$\text{T}_\text{2,short}^{*}$ (ms)} & & \textbf{$\text{T}_\text{1}$ (ms)} & \textbf{$\text{T}_\text{2,long}^{*}$ (ms)} & \textbf{$\text{T}_\text{2,short}^{*}$ (ms)} \\
    \hline
    1 & 0 & 140 & & 58.9 $\pm$ 1.2 & 40.6 $\pm$ 0.9 & 32.9 $\pm$ 1.9 & & 57.4 $\pm$ 2.0 & 52.1 $\pm$ 3.5 & 37.2 $\pm$ 2.7 \\
    2 & 2 & 138 & & 49.8 $\pm$ 2.0 & 32.3 $\pm$ 1.3 & 7.6 $\pm$ 0.8 & & 49.4 $\pm$ 1.9 & 30.5 $\pm$ 2.6 & 10.4 $\pm$ 0.8 \\
    3 & 4 & 135 & & 41.2 $\pm$ 2.9 & 26.5 $\pm$ 1.5 & 6.2 $\pm$ 0.6 & & 42.8 $\pm$ 2.3 & 26.6 $\pm$ 1.4 & 6.1 $\pm$ 0.4 \\
    4 & 4 & 115 & & 40.0 $\pm$ 4.7 & 25.5 $\pm$ 2.2 & 6.2 $\pm$ 0.6 & & 43.0 $\pm$ 1.1 & 26.3 $\pm$ 1.1 & 6.5 $\pm$ 0.7 \\
    5 & 4 & 87 & & 39.4 $\pm$ 2.5 & 27.2 $\pm$ 1.2 & 6.1 $\pm$ 0.4 & & 44.7 $\pm$ 0.6 & 25.9 $\pm$ 0.8 & 6.4 $\pm$ 0.2 \\
    6 & 6 & 132 & & 29.7 $\pm$ 1.9 & 21.9 $\pm$ 1.1 & 6.2 $\pm$ 0.5 & & 38.3 $\pm$ 1.6 & 23.6 $\pm$ 0.8 & 4.5 $\pm$ 0.3 \\
    7 & 8 & 128 & & 28.5 $\pm$ 1.3 & 20.2 $\pm$ 1.2 & 6.3 $\pm$ 0.4 & & 36.8 $\pm$ 0.7 & 23.0 $\pm$ 0.7 & 5.7 $\pm$ 0.3 \\
    \hline
    \end{tabular}}
\end{table*}

\subsection*{Correlation coefficient weighting} 

We investigated the effect of the correlation weighting on the maps by directly evaluating both phantom and brain \na MRF maps weighted with the maximum correlation coefficient only (k = 1) though the k = 1000 highest correlation coefficients. 

To better choose the number of correlation coefficients to apply as a weighting factor in the final maps, we devised a method of selection using limits based on our RM and \textit{a priori} information. First, for each relaxation parameter, a 2D map was generated using \na MRF for every level of unweighted correlation (k = {1,2, ..., 1000}). For the phantom experiments, a range of $\text{T}_\text{1}$, $\text{T}_\text{2,long}^{*}$ and $\text{T}_\text{2,short}^{*}$ determined by the RM were used as limits. For the brain data range, values from the literature for $\text{T}_\text{2,long}^{*}$ in CSF, GM and WM, and $\text{T}_\text{2,short}^{*}$ for GM and WM were used \cite{Kratzer2021, Lommen2018, Ridley2018, Blunck2018, Niesporek2017, Nagel2011, Fleysher2009}. We chose to omit $\text{T}_\text{1}$ in brain from this analysis due to the lack of \na $\text{T}_\text{1}$ values reported at 7 T in the literature. 

These limits were then applied to the maps to generate a set of indices providing the location of pixels where the value of the masked \na MRF mapped parameter fell within the range of the reference. This index set was then used to create a binary mask which was applied back to the original \na MRF 2D maps. These results were plotted as the maximum number of matches made to the subset of pixels within the reference range versus the number k included in matching. 

\subsection*{Statistical Analysis}

We used the two-sided Wilcoxon rank-sum test (WRST) \cite{Riffenberg2006} in MATLAB to compare the values in the ROIs of the 7-compartment phantom, mapped using the RM against \na MRF. Similarly, we used the same test to examine the sensitivity of \na MRF for distinguishing between CSF, GM and WM in brain.

\section{Results}

\subsection*{Phantom}
 
Figure \ref{PhantomFig} shows a diagram of the 7-compartment phantom and the maps from \na MRF: $\text{T}_\text{1}$, $\text{T}_\text{2,long}^{*}$, $\text{T}_\text{2,short}^{*}$, SD, $\Delta\text{B}_\text{1}^{+}$ factor, $\Delta\text{f}_\text{0}$, as well as a map of the average correlation coefficient associated with the signals matched in each ROI (average from 20 correlation coefficients). The ROIs are numbered in correspondence with the data given in Table \ref{phantomTable}, which lists the mean relaxation times calculated for each individual ROI, with their respective standard deviations (STDV), measured using \na MRF and with RM.  

\begin{figure}[t]
    \centering
    \includegraphics[width=0.47\textwidth]{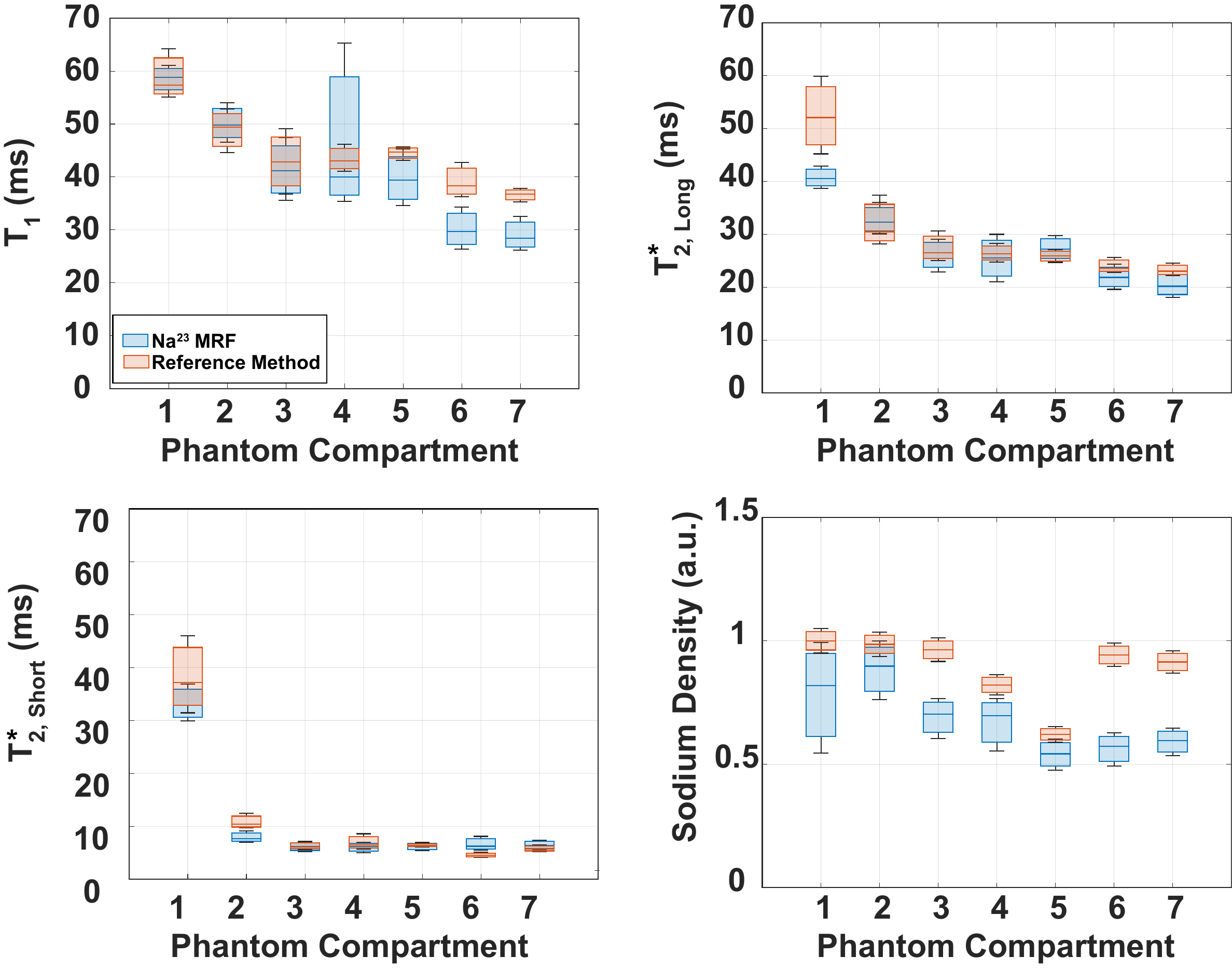}  
    \caption{\textbf{Boxplots of $\text{T}_\text{1}$, $\text{T}_\text{2,long}^{*}$, $\text{T}_\text{2,short}^{*}$ and SD in the phantom compartments: \na MRF vs. reference method (RM).} Relaxation time data corresponds to data listed in Table \ref{phantomTable}. SD calculated using \na MRF was compared to the ground truth (mean value $\pm$5\%) (see Figure \ref{PhantomFig}). Blue boxes represent data from our \na MRF method. Red boxes represent data from RM for relaxation times, and from ground truth for SD.}
    \label{BoxPlot_Phantom}
    \vspace{-10pt}
\end{figure}

Average relaxation times listed in Table \ref{phantomTable} for \na MRF measurements can be easily estimated by visual inspection of the relaxation maps in Figure \ref{PhantomFig}. For comparison, RM values are also presented in Table \ref{phantomTable} and RM maps are shown in Supplementary Figure S4. 

The center ROI 1 contained a solution of 140 mM NaCl and showed slight variations in uniformity for $\text{T}_\text{1}$, $\text{T}_\text{2,long}^{*}$, $\text{T}_\text{2,short}^{*}$ and SD maps, but appeared uniform in maps of $\Delta\text{B}_\text{1}^{+}$ factor and $\Delta\text{f}_\text{0}$. The mean $\text{T}_\text{1}$ for \na MRF was 58.9 $\pm$ 1.2 ms versus 57.4 $\pm$ 2.0 ms for RM. The mean $\text{T}_\text{2,long}^{*}$ for \na MRF was shorter than $\text{T}_\text{1}$ for \na MRF by 18.2 $\pm$ 2.4 ms. For RM, the mean $\text{T}_\text{2,long}^{*}$ (53.1 $\pm$ 3.5 ms) was similar to the mean $\text{T}_\text{1}$ (57.4 $\pm$ 2.0 ms). The difference between mean $\text{T}_\text{2,long}^{*}$ for \na MRF and RM was 11.5 $\pm$ 3.6 ms, where RM was higher. Mean $\text{T}_\text{2,short}^{*}$ values for both \na MRF and RM were lower than $\text{T}_\text{1}$ and $\text{T}_\text{2,long}^{*}$. The difference between $\text{T}_\text{1}$ and $\text{T}_\text{2,short}^{*}$ in \na MRF was  26 $\pm$ 2.2 ms while the difference between $\text{T}_\text{2,long}^{*}$ and $\text{T}_\text{2,short}^{*}$ was 7.7 $\pm$ 2.1 ms. For RM, the difference between  $\text{T}_\text{1}$ and $\text{T}_\text{2,short}^{*}$ was 20.2 $\pm$ 3.4 ms and the difference between $\text{T}_\text{2,long}^{*}$ and $\text{T}_\text{2,short}^{*}$ was 14.9 $\pm$ 4.4 ms.

Loss of uniformity in the phantom compartments was seen in the $\text{T}_\text{1}$ maps of ROI 3 (41.2 $\pm$ 2.9 ms), ROI 4 (40.0 $\pm$ 4.7 ms) and ROI 5 (39.4 $\pm$ 2.5 ms). Out of these ROIs, the most notable artifact appeared in the $\text{T}_\text{1}$ map for ROI 4. In this case, we can align the artifact in the $\text{T}_\text{1}$ map with perturbations in the maps of $\Delta\text{B}_\text{1}^{+}$ factor and $\Delta\text{f}_\text{0}$. The variation noted for the $\text{T}_\text{1}$ map in ROI 4 could also be outlined in the maps for $\text{T}_\text{2,long}^{*}$ and $\text{T}_\text{2,short}^{*}$. Mean $\text{T}_\text{2,long}^{*}$ in ROI 4 was 25.5 $\pm$ 2.2 ms, which represented the highest STDV for $\text{T}_\text{2,long}^{*}$ out of all ROIs. Mean $\text{T}_\text{2,short}^{*}$ in ROI 4 was 6.2 $\pm$ 0.6 ms.   

Relaxation time data is also presented as boxplots in Figure \ref{BoxPlot_Phantom} for comparing \na MRF with RM. SD data is shown in the same figure as a comparison between \na MRF measurements and ground truth (GT), presented as mean value $\pm$5\% STDV.

The boxplots for $\text{T}_\text{1}$ overlap within the interquartile range for ROIs 1 through 5 of the 7-compartment phantom. For ROI 6, the maximum $\text{T}_\text{1}$ for \textsuperscript{23}Na MRF falls 1.93 ms below the minimum for the RM $\text{T}_\text{1}$. In ROI 7 the $\text{T}_\text{1}$ for \na MRF was 2.82 ms below the minimum for RM $\text{T}_\text{1}$. A non-statistically significant difference between RM and \na MRF was found only in ROI 2 with WRST (p = 0.0526). 

The boxplot for $\text{T}_\text{2,long}^{*}$ for ROI 1 shows that the maximum \na MRF $\text{T}_\text{2,long}^{*}$ is 2.31 ms below the minimum RM $\text{T}_\text{2,long}^{*}$. For ROIs 2 through 7, there is overlap within the interquartile ranges between RM and \na MRF. However, a non-statistically significant difference between RM and \na MRF was only found for ROI 4 with WRST (p = 0.0930).  

Interquartile overlap was noted between \na MRF and RM for $\text{T}_\text{2,short}^{*}$ in ROIs 1, 3, 4, 5 and 7. The WRST analysis for these ROIs supported non-statistically significant difference between \na MRF and RM in ROIs 4, 5 and 7 with p = 0.7919, 0.0802 and 0.9054, respectively. In ROI 2, the RM minimum was $\sim$0.6 ms greater than the maximum $\text{T}_\text{2,short}^{*}$ from \na MRF. Finally, for ROI 6, the minimum $\text{T}_\text{2,short}^{*}$ for \na MRF was $\sim$0.6 ms greater than the maximum for RM.

SD of the GT was consistently higher than SD from \na MRF. Overlap in the distributions occurred only between the 2\textsuperscript{nd} quartile of the GT and 4\textsuperscript{th} quartile of \na MRF in ROI 1, between the 2\textsuperscript{nd} quartile of GT and 3\textsuperscript{rd} quartile of \na MRF in ROI 2, and between the 1\textsuperscript{st} quartile of GT and 4\textsuperscript{th} quartile of \na MRF in ROI 5. 

Supplementary Figure S5 shows scatter plots with linear fits between \na MRF and RM for $\text{T}_\text{1}$, $\text{T}_\text{2,long}^{*}$, $\text{T}_\text{2,short}^{*}$, and between \na MRF and GT (mean value $\pm$5\% STDV) for SD, using data from Table \ref{phantomTable}.

\subsection*{Brain}

\begin{figure*}[t!]
    \centering
    \includegraphics[width=\textwidth]{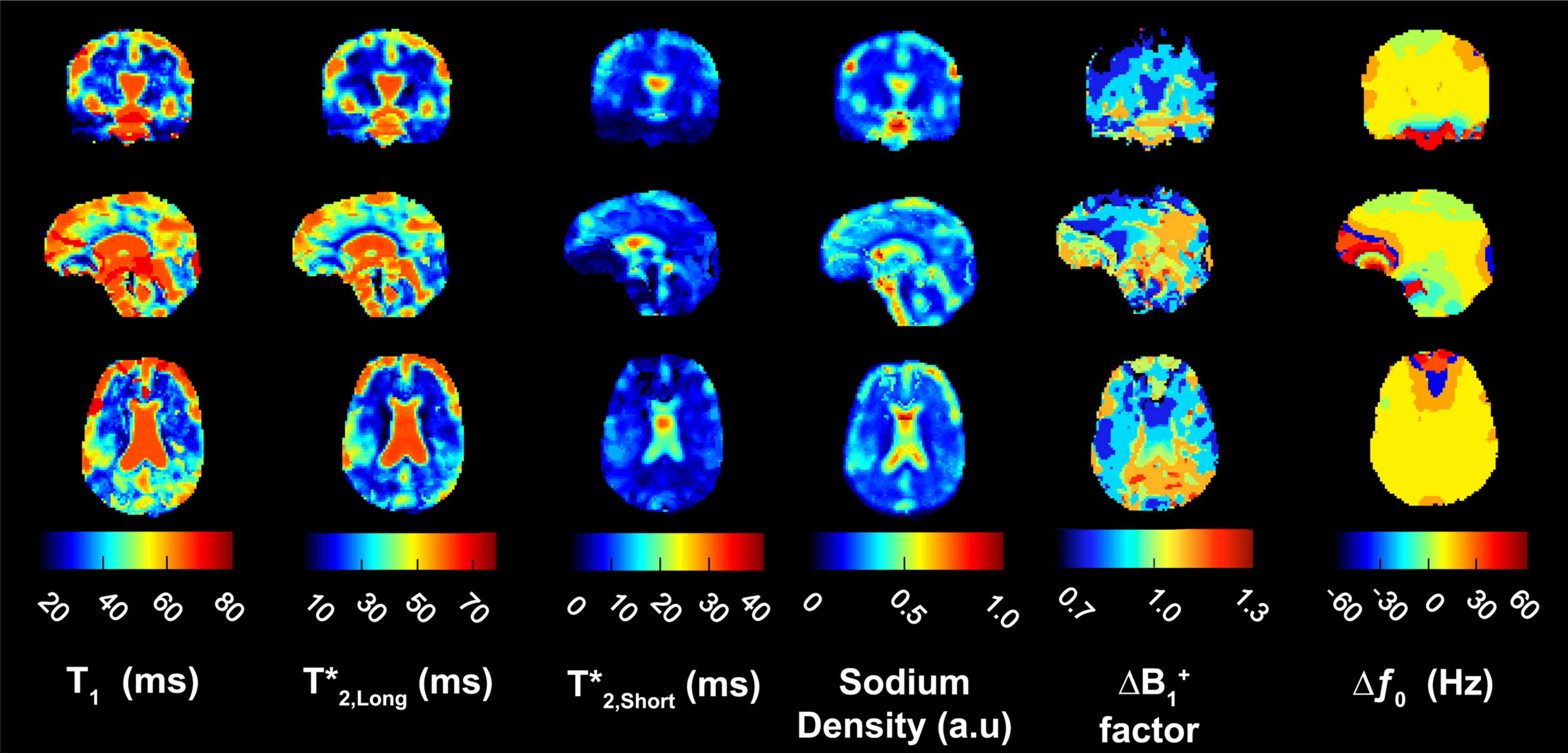}
    \caption{\textbf{Brain maps from \na MRF.} Examples of coronal, sagittal and axial slices for \na relaxation times, SD, $\Delta\text{B}_\text{1}^{+}$ factor and $\Delta\text{f}_\text{0}$ from volunteer 5. All maps were calculated with a correlation coefficient weighting of k = 20.}
    \label{vol5_3D}
\end{figure*}

Figure \ref{vol5_3D} shows the maps for $\text{T}_\text{1}$, $\text{T}_\text{2,long}^{*}$, $\text{T}_\text{2,short}^{*}$, SD, $\Delta\text{B}_{1}^{+}$ and $\Delta\text{f}_\text{0}$ from volunteer 5 in coronal, sagittal and axial slices. We used a weighting factor of k = 20 correlation coefficients during the PC matching process for all volunteers. The most apparent feature in the relaxation times and SD maps was the CSF-filled central ventricles. Contrast from the long $\text{T}_\text{1}$ and $\text{T}_\text{2,long}^{*}$ values in CSF dominate the central ventricular structure in all 3 planes. Contrast from CSF can also be seen within subarachnoid spaces and the cavity along the gyri of the cerebral cortex. In the sagittal slice, we identified the third and fourth ventricles and the occipital horn of the lateral ventricle in maps for $\text{T}_\text{1}$, $\text{T}_\text{2,long}^{*}$, $\text{T}_\text{2,short}^{*}$ and SD. The overall contrast for $\text{T}_\text{2,short}^{*}$ in CSF was reduced compared to $\text{T}_\text{1}$ and $\text{T}_\text{2,long}^{*}$. Here the highest values for  $\text{T}_\text{2,short}^{*}$ were concentrated in the center of the central ventricle in each cross-section. Mean $\text{T}_\text{2,short}^{*}$ values reported in Table \ref{table_volROI} were the result of this unequal distribution of $\text{T}_\text{2,short}^{*}$ over the collective CSF ROI. Mean $\text{T}_\text{2,short}^{*}$ at the center of the central ventricle across all 5 volunteers was 40.1 $\pm$ 0.1 ms.

The normalized SD maps demonstrated high sodium density within CSF and lower sodium density in GM and WM. For $\Delta\text{f}_\text{0}$, extremely negative and positive shifts were mapped around the medial frontal gyrus, shown in the axial position, over the frontal lobe above the nasal sinus cavity, shown in the sagittal plane, and at the base of the medulla near the posterior cerebellum, shown in the coronal plane.
 
\begin{figure*}[t!]
    \centering
    \includegraphics[width=0.95\textwidth]{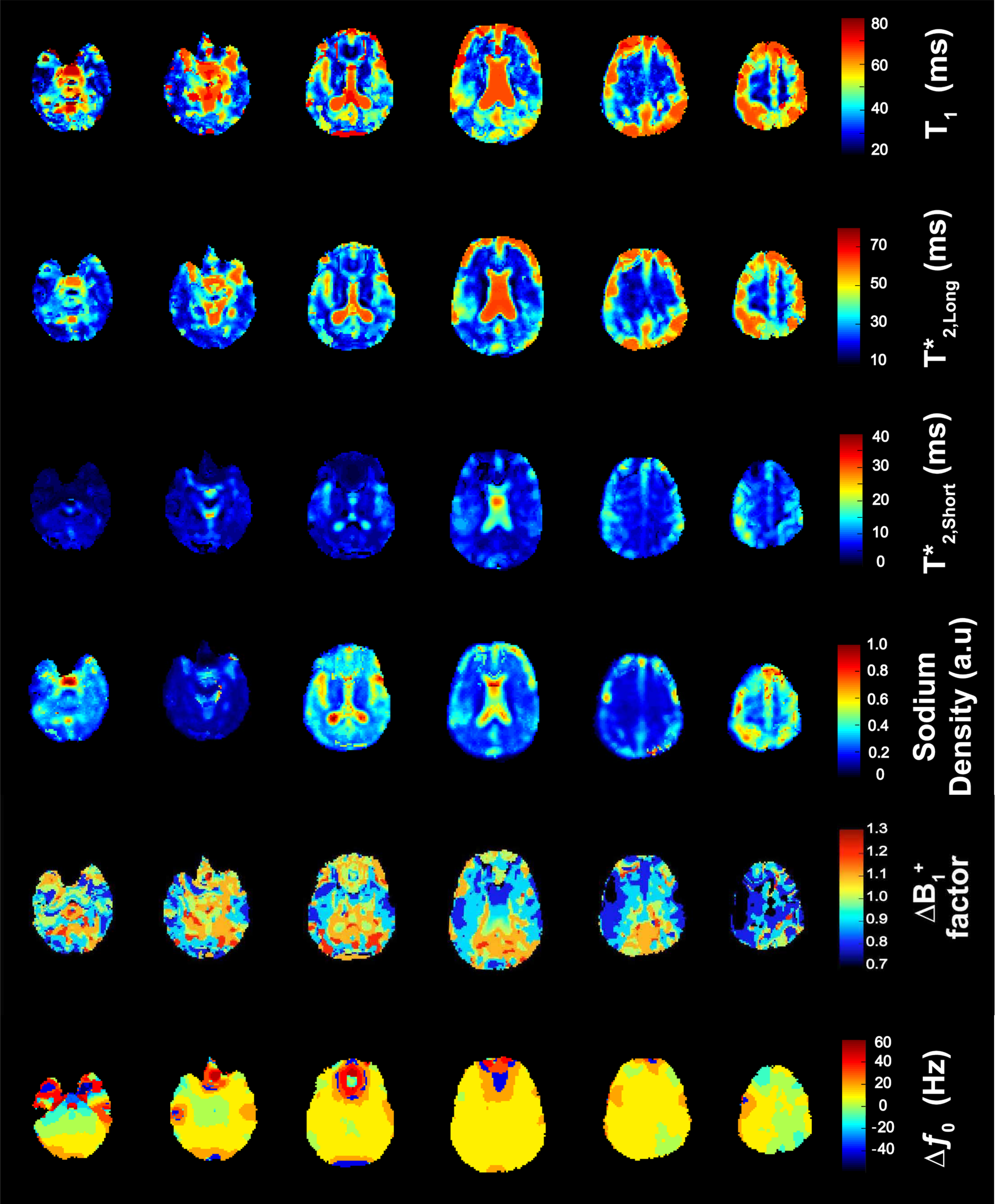}
    \caption{\textbf{Examples of six equally-spaced axial slices of brain maps from \na MRF.} Maps shown are \na relaxation times, SD, $\Delta\text{B}_\text{1}^{+}$ factor and $\Delta\text{f}_\text{0}$ from volunteer 5. All maps were calculated with a correlation coefficient weighting of k = 20.}
    \label{brainSlicesArray}
\end{figure*}

Figure \ref{brainSlicesArray} presents the \na MRF maps in six equidistant axial slices for volunteer 5 with a weighting factor of k = 20 correlation coefficients.

Table \ref{table_volROI} lists the mean \na relaxation times and SD with their respective STDVs in CSF, WM and GM for all 5 subjects. The bottom row lists the mean values with corresponding STDV for each tissue across all five volunteers. The greatest deviation from mean relaxation time across volunteers was recorded in CSF. The shortest $\text{T}_\text{1}$ was measured in volunteer 1 along with the highest STDV (50.7 $\pm$ 15.1 ms) while the longest $\text{T}_\text{1}$ and lowest STDV were recorded in volunteer 5 (63.1 $\pm$ 5.7 ms). The wide range of these values contributed to an overall 8.6\% STDV amongst volunteers. This trend was also observed for $\text{T}_\text{2,long}^{*}$ in CSF, where volunteer 1 contributed 41.7 $\pm$ 16.3 ms and volunteer 5 contributed 57.3 $\pm$ 7.6 ms, toward a mean $\text{T}_\text{2,long}^{*}$ of 49.7 $\pm$ 6.3 ms across all five volunteers. For $\text{T}_\text{2,short}^{*}$ in CSF, a mean value of 12.5 $\pm$ 3.0 ms was measured over the five volunteers with \na MRF. Finally, a good agreement between relaxation times in GM and WM was noted across volunteers, as indicated by the low corresponding STDVs.

\begin{figure}[t!]
    \centering
    \includegraphics[width=0.47\textwidth]{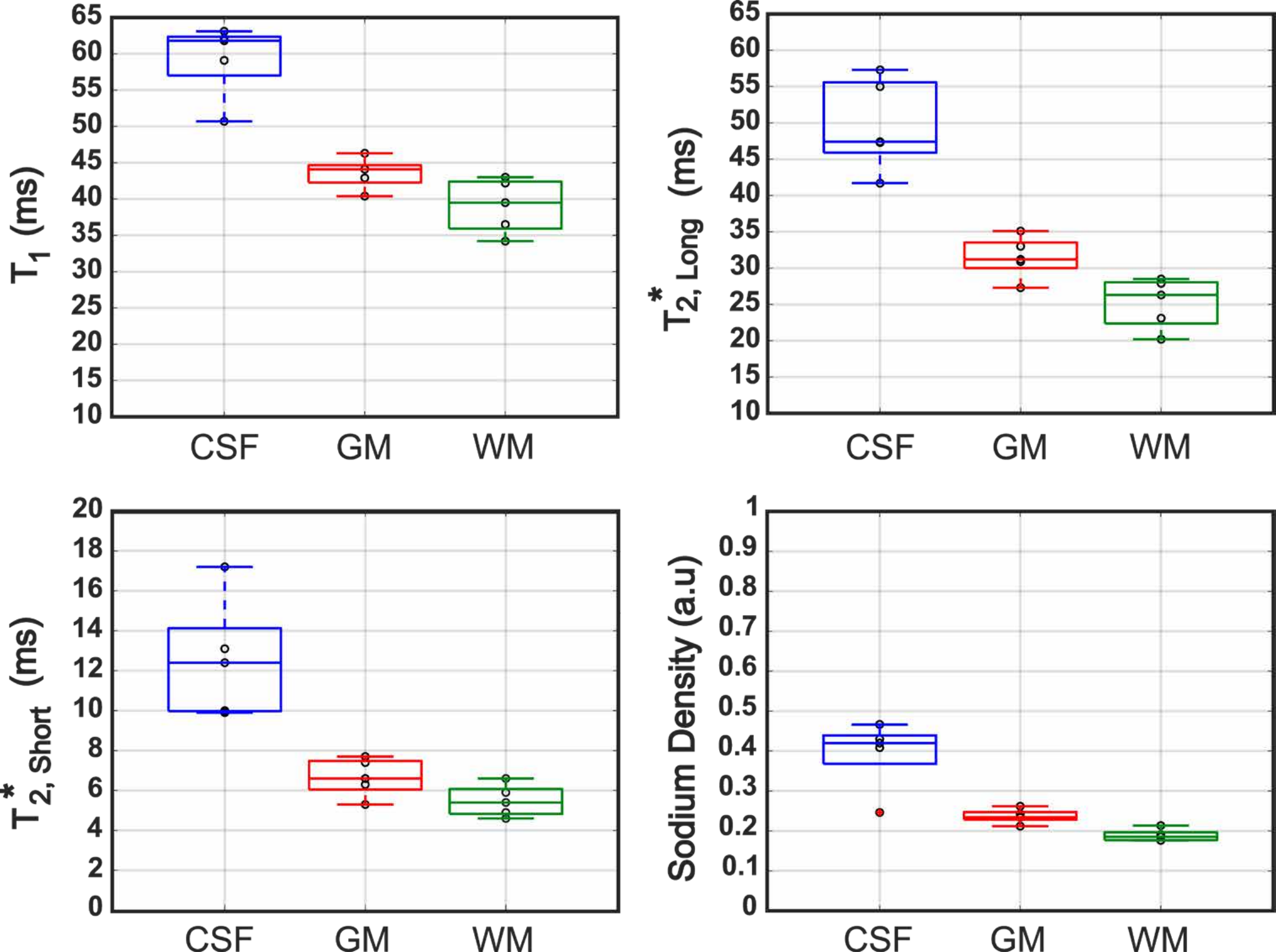}
    \caption{\textbf{Boxplots of the mean $\text{T}_\text{1}$, $\text{T}_\text{2,long}^{*}$, $\text{T}_\text{2,short}^{*}$ and SD in CSF, GM and WM measured in the brain of 5 volunteers.} Mean values were measured on the center slice of each of the three planes (see Table \ref{table_volROI}). Data points for individual volunteers are indicated with black circles, and outliers are marked in red. The lines inside each box represent the median values. \textit{Abbreviations:} GM, grey matter; WM, white matter; CSF, cerebrospinal fluid.}
    \label{BoxPlot_Brain}
\end{figure}

Figure \ref{BoxPlot_Brain} summarizes the results from Table \ref{table_volROI} into boxplots. We found that the median $\text{T}_\text{1}$ in CSF was 61.8 ms and that there was no overlap in $\text{T}_\text{1}$ ranges between either the GM or WM groups. Strong differentiation between CSF and GM/WM was corroborated statistically by WRST (p = 0.0079). The median $\text{T}_\text{1}$ for GM was 44.1 ms and the median $\text{T}_\text{1}$ for WM was 39.5 ms. The lower adjacent value in GM (40.4 ms) was positioned below the third quartile of WM with the upper adjacent value in WM (43 ms) positioned above the first quartile of GM. Despite the overlap in these regions, a statistically significant difference between $\text{T}_\text{1}$ in GM and WM was observed (p = 0.0476). 

For $\text{T}_\text{2,long}^{*}$ in CSF, the median value was 47.4 ms. Again, there was no overlap of $\text{T}_\text{2,long}^{*}$ for CSF with $\text{T}_\text{2,long}^{*}$ in either GM or WM. The differences were statistically significant (p = 0.0079) between GM and CSF, and between WM and CSF. The median $\text{T}_\text{2,long}^{*}$ in GM was 31.2 ms with a lower adjacent value of 27.3 ms, which fell below the third quartile of WM. The median $\text{T}_\text{2,long}^{*}$ in WM was 26.3 ms. WRST comparison between $\text{T}_\text{2,long}^{*}$ in GM and WM indicated a statistically significant difference (p = 0.0317).

Median $\text{T}_\text{2,short}^{*}$ in CSF, GM and WM were 12.4 ms, 6.6 ms and 5.4 ms, respectively. As was the case in $\text{T}_\text{1}$ and $\text{T}_\text{2,long}^{*}$, the range of $\text{T}_\text{2,short}^{*}$ shown for the boxplot of CSF in Figure \ref{BoxPlot_Brain} did not align within the ranges of $\text{T}_\text{2,short}^{*}$ in GM or WM. Similarly to statistical results results for $\text{T}_\text{1}$ and $\text{T}_\text{2,long}^{*}$, $\text{T}_\text{2,short}^{*}$ in CSF was statistically significantly different from $\text{T}_\text{2,short}^{*}$ in both GM and WM (p = 0.0079). There was no statistically significant difference in $\text{T}_\text{2,short}^{*}$ between GM and WM (p = 0.1111). This was corroborated by the respective boxplots in Figure \ref{BoxPlot_Brain} where we noted that the upper adjacent of WM was equal to the median of GM (6.6 ms).

For normalized SD in CSF, we found a maximum of 0.467, a median of 0.420 and a minimum of 0.0246, which was considered an outlier. The lower adjacent of CSF was 0.368 and did not overlap with any data grouped for either GM or WM. The minimum value recorded for GM was 0.212, which was nearly equivalent to the maximum of WM (0.213). A statistically significant difference was observed for SD between GM and WM (p = 0.0159). 

Table \ref{litVals} provides a comparison between the relaxation times obtained using our \na MRF method and those reported in the literature at 7 T. For $\text{T}_\text{1}$ in CSF measured using our \na MRF method, the mean value (59.4 $\pm$ 5.1 ms) was comparable to the mean $\text{T}_\text{1}$ reported by Kratzer et al. \cite{Kratzer2021} (61.9 $\pm$ 2.8 ms) using their \na MRF method. Although not included in the Table \ref{litVals}, the $\text{T}_\text{1}$ for WM reported using our \na MRF was within the range reported by Coste et al. \cite{Coste2019} for WM $\text{T}_\text{1}$ at 3 T. $\text{T}_\text{2}^{*}$ values for CSF were reported throughout the literature as monoexponential fits with values ranging from 46.3 $\pm$ 6.3 ms \cite{Kratzer2021} to 57.2 $\pm$ 6.6 ms \cite{Blunck2018}. This range indicated that our measured value of 49.7 $\pm$ 6.3 ms for $\text{T}_\text{2,long}^{*}$ was in good agreement with the literature. Our mean $\text{T}_\text{2,short}^{*}$ of 12.5 $\pm$ 3.0 ms, however, was out of range compared to these same values.

We did not find values for $\text{T}_\text{1}$ specific to GM and WM during our literature search (which we limited to data recorded at 7 T), so we compared our results to the $\text{T}_\text{1}$ for unspecified brain tissue determined by Kratzer et al.\cite{Kratzer2021}. In this case, our $\text{T}_\text{1}$ measurement for WM was in good agreement with their value for brain tissue, but our value for GM was $\sim$11 ms longer. 

For $\text{T}_\text{2,long}^{*}$ in GM, our measured value of  31.5 $\pm$ 2.9 ms was in agreement with previous reports \cite{Blunck2018,Ridley2018,Niesporek2017,Fleysher2009}, while our measured value of 6.7 $\pm$ 1.0 ms for $\text{T}_\text{2,short}^{*}$ was similar to that found by Ridley et al. \cite{Ridley2018}, but slightly longer than other reports \cite{Blunck2018,Niesporek2017}.

For $\text{T}_\text{2,long}^{*}$ in GM, our measured value of  31.5 $\pm$ 2.9 ms was similar to the value determined by Ridley et al. \cite{Ridley2018} and Fleysher et al. \cite{Fleysher2009} and within the deviation of average values reported by Blunck et al. \cite{Blunck2018} and Niesporek et al. \cite{Niesporek2017}, respectively. Our measured value of 6.7 $\pm$ 1.0 ms for $\text{T}_\text{2,short}^{*}$ was similar to that found by Ridley et al. (5.0 $\pm$0.9) \cite{Ridley2018}, but slightly longer than those values reported by Blunck et al. \cite{Blunck2018} and Niesporek et al. \cite{Niesporek2017}. For $\text{T}_\text{2,long}^{*}$ in WM, we calculated a mean of 25.2 $\pm$ 3.5 ms, which was within the range of 22.4 $\pm$ 7.8 ms \cite{Blunck2018} to 40.0 $\pm$ 5.2 ms \cite{Ridley2018}  reported for $\text{T}_\text{2,long}^{*}$ in the literature and provided in Table \ref{litVals}. In this case, our value was similar to those reported by Blunck et al. \cite{Blunck2018} and Niesporek et al. \cite{Niesporek2017}, but lower than those reported by Lommen et al.\cite{Lommen2018}, Ridley et al. \cite{Ridley2018} and Fleysher et al. \cite{Fleysher2009}, respectively. Our WM $\text{T}_\text{2,short}^{*}$ value of 5.5 $\pm$ 0.8 ms our values were similar to that reported by Lommen et al. \cite{Lommen2018}, within the deviation of the reported value from Ridley et al.\cite{Ridley2018} but higher than those reported by Blunck et al. \cite{Blunck2018} and Niesporek et al.\cite{Niesporek2017}, respectively. Finally, both the $\text{T}_\text{2,long}^{*}$ and $\text{T}_\text{2,short}^{*}$ values that we measured in GM and WM were in good agreement with values reported for brain tissue by Kratzer et al. \cite{Kratzer2021}, which the authors provided as GM and WM combined.

\subsection*{Correlation coefficient weighting}

Supplementary Figure S6 shows the maps of the 7-compartment phantom produced after matching was performed using different numbers of correlation coefficients as weighting factors, from the maximum correlation only (k = 1) through k = 1000.

Supplementary Figure S7 shows the maps for volunteer 5 representing brain results produced after matching was performed using different numbers of correlation coefficients as weighting factors, from the maximum correlation only (k = 1) through k = 1000.

Supplementary Figure S8 summarizes the results of matching in the 7-compartment phantom with different k values against the RM in a series of boxplots. When k = 20, there was more overlap between the interquartile regions of RM and $\text{T}_\text{1}$ as compared to the single maximum valued correlation, and the most overlap in interquartile regions for all ROIs in $\text{T}_\text{2,long}^{*}$. Although STDV was reduced for $\text{T}_\text{1}$ as the number of correlation coefficients increases, there was loss of overlap between RM and \na MRF for $\text{T}_\text{2,long}^{*}$ in ROIs 6 and 7, respectively. Overall, $\text{T}_\text{2,short}^{*}$ did not change significantly between k = 1 and k = 200. 

Supplementary Figure S9 shows a series of graphs for the relaxation times measured in each ROI of the 7-compartment phantom plotting the maximum number of matches made to a subset of pixels within a reference range limited by values from the RM. Examining the trends in the graphs, we noted that the plot most often changed direction in $\text{T}_\text{2,long}^{*}$ from a high number of maximum correlations per pixel match to a global decrease followed by leveling off in the region between k = 10 to k = 50. Direction changes were also noted in $\text{T}_\text{2,short}^{*}$ graphs in the neighborhood of k = 100.  

Supplementary Figure S10 shows similar graphs for $\text{T}_\text{2,long}^{*}$ in CSF, GM and WM and $\text{T}_\text{2,short}^{*}$ in GM and WM for each of the 5 volunteers. The maximum number of matches made to a subset of pixels within each ROI were limited by ranges defined by literature values taken from Table \ref{litVals}. While graphs of $\text{T}_\text{2,short}^{*}$ do not show any notable trend, there are direction changes in the graphs of $\text{T}_\text{2,long}^{*}$ in the neighborhood of k = 20 for CSF, GM and WM.

\section{Discussion}

In this work we demonstrated a refined approach to quantitative mapping of $\text{T}_\text{1}$, $\text{T}_\text{2,long}^{*}$, $\text{T}_\text{2,short}^{*}$ and SD using \textsuperscript{23}Na MRF with correlation coefficient weighting. We constructed a comprehensive dictionary that included combinations of $\text{T}_\text{1}$, $\text{T}_\text{2,long}^{*}$, $\text{T}_\text{2,short}^{*}$, $\Delta\text{B}_\text{1}^{+}$ factor and $\Delta\text{f}_\text{0}$. We implemented a 3D FLORET sequence with an optimized 23-pulse variable FA/PA MRF train capable of full brain coverage in about 30 min without varying TE or delays between the pulses in the \na MRF pulse train. While the ISTO spin system simulation and PC matching procedure were based on our previous work \cite{Gilles2017}, this study integrated a new \na MRF pulse train optimization protocol and refined matching criteria. Our method was tested in a 7-compartment phantom, and successfully applied for brain mapping in five healthy volunteers at 7 T.  

Because $\Delta\text{B}_\text{1}^{+}$ and $\Delta\text{f}_\text{0}$ were included as dictionary parameters, the spatial influences of transmit inhomogeneity and frequency shift on relaxation times were partially accounted for in the matching process. RF field variations were most pronounced in the phantom data (Figure \ref{PhantomFig}), due to its high average relative permittivity. On the other hand, $\Delta\text{f}_\text{0}$ variations were more pronounced in vivo, due to air-filled structures such as the ear canal and maxillary sinus cavities (Figure \ref{brainSlicesArray}). Some of the artifacts noted in the relaxation maps could be lined up with similar artifacts in maps of $\Delta\text{B}_\text{1}^{+}$ and $\Delta\text{f}_\text{0}$. This alignment of artifacts demonstrated how the spatial influence of transmit inhomogeneity and frequency shift were articulated in the relaxation maps through the matching process, and how inhomogeneity influenced the matching of signals in the relaxation maps. These results indicated that broadening the ranges for $\Delta\text{B}_\text{1}^{+}$ and $\Delta\text{f}_\text{0}$ in the dictionary, or reducing the step size for these entries, might improve the appearance of these artifacts in the parameter maps. For this study, our choice of $\Delta\text{B}_\text{1}^{+}$ factor and $\Delta\text{f}_\text{0}$ steps reflect a compromise between computation time and the possible range for $\Delta\text{B}_\text{1}^{+}$ factor and $\Delta\text{f}_\text{0}$ variation in head at 7T. Also significant is that our \na MRF method does not use the maps for $\Delta\text{B}_\text{1}^{+}$ and $\Delta\text{f}_\text{0}$ generated through the matching process as a post-processing correction to the final relaxation maps. All maps produced using our \na MRF technique ($\text{T}_\text{1}$, $\text{T}_\text{2,long}^{*}$, $\text{T}_\text{2,short}^{*}$ and sodium density) were generated from matching the dictionary of simulated signals to real data where $\Delta\text{B}_{1}^{+}$ and $\Delta\text{f}_\text{0}$ corrections were built into the dictionary.

\newpage

\onecolumn

\begin{sidewaystable}

    \caption{\textbf{Mean \na MRF values in brain in 5 volunteers.} Data shown represents mean relaxation times and normalized sodium density (SD) $\pm$ standard deviations (STDV) measured in three brain tissue ROIs (cerebrospinal fluid [CSF], grey matter [GM], white matter [WM]) for each volunteer (VOL). Mean values were calculated over the center slices of the axial, coronal and sagittal planes of the brain. The bottom row of each column corresponds to the mean value $\pm$ STDV of the mean values measured in each ROI across all volunteers \label{table_volROI}}
    \centering
    \resizebox{\textwidth}{!}{
    \scriptsize
    \begin{tabular}{llllllllllllllll}   
    \hline
    \multicolumn{1}{l}{} & & \multicolumn{4}{l}{\textbf{CSF}} & & \multicolumn{4}{l}{\textbf{GM}} & & \multicolumn{4}{l}{\textbf{WM}} \\ 
    \cline{3-6} \cline{8-11} \cline{13-16} 
    \textbf{VOL} & & \textbf{$\text{T}_\text{1}$ (ms)} & \textbf{$\text{T}_\text{2,long}^{*}$ (ms)} & \textbf{$\text{T}_\text{2,short}^{*}$ (ms)} & \textbf{SD} & & \textbf{$\text{T}_\text{1}$ (ms)} & \textbf{$\text{T}_\text{2,long}^{*}$ (ms)} & \textbf{$\text{T}_\text{2,short}^{*}$ (ms)} & \textbf{SD} & & \textbf{$\text{T}_\text{1}$ (ms)} & \textbf{$\text{T}_\text{2,long}^{*}$ (ms)} & \textbf{$\text{T}_\text{2,short}^{*}$ (ms)} & \textbf{SD}\\
    \hline
    1 & & 50.7 $\pm$ 15.1 & 41.7 $\pm$ 16.3 & 10.0 $\pm$ 8.8 & 0.246 $\pm$ 0.134 & & 44.1 $\pm$ 12.5 & 33.0 $\pm$ 11.2 & 6.6 $\pm$ 3.6 & 0.212 $\pm$ 0.072 & & 43.0 $\pm$ 13.6 & 27.9 $\pm$ 12.1 & 6.6 $\pm$ 3.9 & 0.177 $\pm$ 0.054 \\
    2 & & 62.1 $\pm$ 7.0 & 55.0 $\pm$ 9.6 & 17.2 $\pm$ 9.4 & 0.467 $\pm$ 0.172 & & 44.1 $\pm$ 12.0 & 31.2 $\pm$ 9.2 & 7.4 $\pm$ 3.8 & 0.234 $\pm$ 0.067 & & 36.5 $\pm$ 10.5 & 23.1 $\pm$ 8.5 & 5.4 $\pm$ 2.9 & 0.191 $\pm$ 0.063 \\
    3 & & 61.8 $\pm$ 10.2 & 47.3 $\pm$ 13.9 & 12.4 $\pm$ 7.8 & 0.420 $\pm$ 0.169 & & 42.9 $\pm$ 12.5 & 30.9 $\pm$ 9.2 & 5.3 $\pm$ 2.7 & 0.262 $\pm$ 0.711 & & 39.5 $\pm$ 13.3 & 26.3 $\pm$ 9 .8 & 4.6 $\pm$ 3.0 & 0.213 $\pm$ 0.061 \\
    4 & & 59.1 $\pm$ 10.3 & 47.4 $\pm$ 12.9 & 9.9 $\pm$ 7.2 & 0.410 $\pm$ 0.165 & & 40.4 $\pm$ 12.8 & 27.3 $\pm$ 8.3 & 6.3 $\pm$ 3.1 & 0.234 $\pm$ 0.060 & & 34.2 $\pm$ 11.8 & 20.2 $\pm$ 9.7 & 4.9 $\pm$ 2.6 & 0.176 $\pm$ 0.062 \\
    5 & & 63.1 $\pm$ 5.7 & 57.3 $\pm$ 7.6 & 13.1 $\pm$ 7.1 & 0.430 $\pm$ 0.128 & & 46.3 $\pm$ 12.9 & 35.1 $\pm$ 11.1 & 7.7 $\pm$ 3.4 & 0.242 $\pm$ 0.071 & & 42.2 $\pm$ 11.0 & 28.5 $\pm$ 12.2 & 5.9 $\pm$ 2.7 & 0.185 $\pm$ 0.061 \\
    \hline
    Mean & & 59.4 $\pm$ 5.1 & 49.7 $\pm$ 6.3 & 12.5 $\pm$ 3.0 & 0.395 $\pm$ 0.086 & & 43.6 $\pm$ 2.1 & 31.5 $\pm$ 2.9 & 6.7 $\pm$ 1.0 & 0.237 $\pm$ 0.018 & & 39.1 $\pm$ 3.7 & 25.2 $\pm$ 3.5 & 5.5 $\pm$ 0.8 & 0.188 $\pm$ 0.151 \\
    \hline
    \end{tabular}}
    
    \bigskip \bigskip \bigskip \bigskip \bigskip \bigskip \bigskip \bigskip \bigskip

    \caption{\textbf{Comparison of the relaxation times measured in brain at 7 T between the correlation-weighted (CW)\na MRF method and values from the literature.} Relaxation times values were compared in cerebrospinal fluid (CSF), grey matter (GM) and  white matter (WM). "Brain Tissue" was included where a distinction was not made between GM and WM.\label{litVals}}
    \centering
    \resizebox{\textwidth}{!}{
    \scriptsize
    \begin{tabular}{llllllllllllllllll}
    \hline
    \multicolumn{2}{l}{} & & \multicolumn{3}{l}{\textbf{CSF}} & & \multicolumn{3}{l}{\textbf{GM}} & & \multicolumn{3}{l}{\textbf{WM}} & & \multicolumn{3}{l}{\textbf{Brain Tissue}} \\ 
    \cline{4-6} \cline{8-10} \cline{12-14} \cline{16-18}
    \newline
    \textbf{Reference} & \textbf{Method} & & \textbf{$\text{T}_\text{1}$ (ms)} & \textbf{$\text{T}_\text{2,long}^{*}$ (ms)} & \textbf{$\text{T}_\text{2,short}^{*}$ (ms)} & & \textbf{$\text{T}_\text{1}$ (ms)} & \textbf{$\text{T}_\text{2,long}^{*}$ (ms)} & \textbf{$\text{T}_\text{2,short}^{*}$ (ms)} & & \textbf{$\text{T}_\text{1}$ (ms)} & \textbf{$\text{T}_\text{2,long}^{*}$ (ms)} &   \textbf{$\text{T}_\text{2,short}^{*}$ (ms)} & & \textbf{$\text{T}_\text{1}$ (ms)} & \textbf{$\text{T}_\text{2,long}^{*}$ (ms)} & \textbf{$\text{T}_\text{2,short}^{*}$ (ms)} \\
    \hline
    Kratzer et al. \cite{Kratzer2021} & \na MRF & & 61.9 $\pm$ 2.8 & 46.3 $\pm$ 4.5 & ~ & & ~ & ~ & ~ & & ~ & ~ & ~ & & 35.0 $\pm$ 3.2 & 29.3 $\pm$ 3.8 & 5.5 $\pm$ 1.3 \\
    Lommen et al. \cite{Lommen2018} & DA-3DPR & & ~ & 53.6 $\pm$ 6.9 & ~ & & ~ & ~ & ~ & & ~ & 35.7 $\pm$ 2.4 & 5.1 $\pm$ 0.8 & & ~ & ~ & ~ \\ 
    ~  & DA-3DPR & & ~ & 54.4 $\pm$ 5.7 & ~ & & ~ & ~ & ~ & & ~ & 34.4 $\pm$ 1.5 & 4.2 $\pm$0.4 & & ~ & ~ & ~ \\
    Ridley \cite{Ridley2018} & DA-3DPR & & ~ & ~ & ~ & & ~ & 33.9 $\pm$ 5.9 & 5.0 $\pm$ 0.9 & & ~ & 34.0 $\pm$ 5.2 & 4.5 $\pm$ 0.6 & & ~ & ~ & ~ \\
    Blunck et al. \cite{Blunck2018} & 3D-MERINA & & ~ & 57.2 $\pm$ 6.6 & ~ & & ~ & 25.9 $\pm$ 8.3 & 2.0 $\pm$ 2.1 & & ~ & 22.4 $\pm$ 7.8 & 2.0 $\pm$ 2.1 & & ~ & ~ & ~\\
    Niesporek \cite{Niesporek2017} & DA-3DPR & & ~ & 46.9 $\pm$ 2.1 & ~ & & ~ & 36.4 $\pm$ 3.1 & 5.4 $\pm$ 0.2 & & ~ & 23.3 $\pm$ 2.6 & 3.5 $\pm$ 0.1 & & ~ & ~ & ~ \\
    Nagel et al. \cite{Nagel2011} & DA-3DPR & & ~ & 56.0 $\pm$ 4.0 & ~ & & ~ & ~ & ~ & & ~ & ~ & ~ & & ~ & ~ & ~\\
    Fleysher et al. \cite{Fleysher2009} & GRE & & ~ & 54.0 $\pm$ 4.0 & ~ & & ~ & 28.0 $\pm$ 2.0 & ~ & & ~ & 29.0 $\pm$ 2.0 & ~ & & ~ & ~ & ~ \\
    \hline
    Our work & CW \na MRF & & 59.4 $\pm$ 5.1 & 49.7 $\pm$ 6.3 & 12.5 $\pm$ 3.0 & & 43.6 $\pm$ 2.1 & 31.5 $\pm$ 2.9 & 6.7 $\pm$ 1.0 & & 39.1 $\pm$ 3.7 & 25.2 $\pm$ 3.5 & 5.5 $\pm$ 0.8 & & ~ & ~ & ~ \\
    \hline
    \end{tabular}}
    
\end{sidewaystable}

\newpage

\twocolumn

The low resolution used to offset the sodium SNR deficits makes precise measurements in isolated tissues difficult. Even at the current resolution, low SNR remains a hurdle. Speckle noise can be seen in some of the maps in Figures \ref{vol5_3D} and \ref{brainSlicesArray}, that overlap in regions where changes in $\Delta\text{B}_\text{1}^{+}$ factor or $\Delta\text{f}_\text{0}$ were apparent, corresponding to areas of low SNR. We first tried to overcome some of these caveats by denoising the images prior to matching. While denoising did offer subtle improvement in image quality by removing some pixels with outlying intensity, the image quality was not enhanced enough to make a difference in the matching process.

Matching with PC was performed voxelwise between the data and the dictionary. We ordered the signal matches for each individual voxel according to their respective correlation coefficients and then produced maps reflecting data that included a subset of matches for each voxel. We later refined this technique by using the correlation coefficient value per voxel signal match as a weighting factor. While the time required for matching and map reconstruction was longer than the fitting times of the RM for $\text{T}_\text{1}$, $\text{T}_\text{2,long}^{*}$ and $\text{T}_\text{2,short}^{*}$, the \na MRF method has the benefit of mapping not only $\text{T}_\text{1}$, $\text{T}_\text{2,long}^{*}$ and $\text{T}_\text{2,short}^{*}$ but also $\Delta\text{B}_{1}^{+}$, $\Delta\text{f}_\text{0}$ and SD in a single acquisition. Furthermore, the combined scan time for RM based on the scan time in the phantom, would be more than 2 h versus a scan time of about 30 min for brain \na MRF.

Changes in the maps of correlation coefficient value averaged over increasing numbers of coefficients were minor. This was because the differences between the maximum value and an average of some subset of coefficients for any single voxel were in the thousandths. Despite such minor changes in value, each correlation represents a potential match between the dictionary and data. We acknowledge that examining correlation in non-convex space and within a noisy environment opens the possibility that some higher valued correlations were calculated for signals representing local minima as opposed to a "true" match. Our choice to include multiple matches based on correlation coeffcient weighting increased the probability that some match resulted from a "true" match as opposed to a local minimum. It may be possible to further improve matching by accounting for non-convexity in the reconstruction process \cite{Haacke1990, Zhao2018, Duarte2020}. 

In the phantom, we weighted the data with 20 coefficients. This reduced the slight inhomogeneity in $\text{T}_\text{1}$ and $\text{T}_\text{2,long}^{*}$ and provided the best agreement between \na MRF and the RM data. While weighting beyond 20 coefficients continued to slightly smooth artifacts and improve the agreement between \na MRF and RM in $\text{T}_\text{1}$, the agreement to RM in $\text{T}_\text{2,long}^{*}$ began to diverge in ROIs 6 and 7 at k = 50. There was no significant change in values beyond the maximum correlation for $\text{T}_\text{2,short}^{*}$. Similarly, for SD, an improved overlap in the boxplots between \na MRF and RM is seen at k = 20 with no change beyond this value. As expected, \na $\text{T}_\text{1}$ decreased with increasing agar concentration consistent with an increasingly restricted environment \cite{Madelin2014}. Note that since B\textsubscript{1}\textsuperscript{+} and B\textsubscript{0} maps were not acquired in the phantom reference methods, this lack of B\textsubscript{1}\textsuperscript{+} and B\textsubscript{0} correction might also partially explain the discrepancy between RM and \na MRF measurements.  

Moreover, while results in phantoms show quite good agreement between our proposed \na MRF method and RM, we acknowledge the difficulty of obtaining accurate relaxation times with \na MRI in general, as most values are within 5 ms here, which is likely within SNR variations between the methods that we compared, since \na imaging is inherently a low-SNR technique, even for RM.

Matching in vivo was also completed using weighting with k = 20 correlation coefficients. In this case, increasing the number of coefficients introduced signals with lower value correlation coefficients for CSF. Interestingly, this had the effect of increasing the values of $\text{T}_\text{2,short}^{*}$ within CSF while slightly decreasing $\text{T}_\text{1}$ and $\text{T}_\text{2,long}^{*}$. This was combined with an overall increase in the $\text{T}_\text{1}$ of brain tissue and decrease in SD. Essentially, weighting signals by their correlation coefficients operated as a smoothing kernel by including a range of highly similar fingerprints. We compared the smoothing effect to that of applying a Gaussian filter prior to matching which is shown in Supplementary Figure S11. Comparison of parameter maps at the maximum correlation with and without the addition of the Gaussian filter demonstrated no effective improvement in the visual appearance for the brain data. Furthermore, this outcome did not change when additional correlations were included. 

Previous studies have indicated that differences in tissue structure and mobility affect the residual quadrupolar interaction arising from \na spins interacting with their environment\cite{Madelin2014}. More restricted environments, such as myelinated white matter, are associated with stronger residual quadrupolar couplings that lead to faster decay components. However, there is also evidence that sodium exists as different pools of nuclei within individual brain tissues, where some nuclei exhibit biexponential relaxation while other spin pool distributions demonstrate monoexponential behavior\cite{Stobbe2016}. Other works support similar considerations that regional differences in structural composition within GM and WM \cite{Ridley2018, Kolbe2020}, or local changes induced within the cellular environment of the GM and WM \cite{Petracca2016} confound the ability to explicitly distinguish between these tissues, respectively. Combined with the inherently low SNR of sodium imaging, these factors make distinguishing GM and WM through relaxation mapping difficult. In evaluating the ability of our \na MRF method to distinguish between GM and WM we compared our results against 7 references (Table \ref{litVals}) where $\text{T}_\text{1}$, $\text{T}_\text{2,long}^{*}$ and $\text{T}_\text{2,short}^{*}$ were reported as mean values across a cohort of volunteers in each study. Using similar mean value analysis, we reported comparable results in the last line of Table \ref{litVals}, which indicated that our method could distinguish between GM and WM as well as other methods listed. Additionally, we were able to demonstrate that the mean $\text{T}_\text{1}$ and $\text{T}_\text{2,long}^{*}$ were different enough to distinguish between GM and WM using statistical analysis, provided in the Results section. Despite the statistical analysis and mean value support for previous findings, visual inspection of the brain maps of the five volunteers alone was not sufficient for differentiating between GM and WM. 

For those references considered in Table \ref{litVals} where $\text{T}_\text{2,long}^{*}$ and $\text{T}_\text{2,short}^{*}$ were reported, there is a wide range between the shortest and longest $\text{T}_\text{2,long}^{*}$ and $\text{T}_\text{2,short}^{*}$ noted between different sources. Different fitting techniques or data acquisition schemes could be the culprit as well as some variation in relaxation time within GM and WM.

Examining the references in Table \ref{litVals}, we noted that several acquisition schemes were in play by different authors; Blunck et al. \cite{Blunck2018}., Lommen et al. \cite{Lommen2018}, Ridley et al. \cite{Ridley2018}, Niesporek et al. \cite{Niesporek2017} and Nagel et al. \cite{Nagel2011} applied variations of multiple echo acquisitions with varied TE and 3D radial readouts. Fleysher et al. \cite{Fleysher2009} used a gradient recalled echo (GRE) sequence with a Cartesian readout. In these works, the authors determined $\text{T}_\text{2,long}^{*}$ and $\text{T}_\text{2,short}^{*}$ using using voxel wise fitting in image space where a signal model was defined as biexponential decay and the signal contributions from satellite and central transitions were expected to demonstrate a $\text{T}_\text{2,short}^{*}$:$\text{T}_\text{2,long}^{*}$ ratio of 0.6:0.4. In some work, this ratio was explicitly built-in to the fitting model itself\cite{Blunck2018,Lommen2018,Niesporek2017,Fleysher2009}. Kratzer and colleagues have previously reported two methods for \na MRF. The first involved using modified Bloch equations as a simplified simulation for the \na spin $\frac{3}{2}$ system, was capable of single slice 2D acquisition and suffered from low efficiency \cite{Kratzer2020}. Their followup to this work utilized an ISTO simulation technique and was expanded to 3D \cite{Kratzer2021}. In our study we also employed the ISTO formalism to model \na relaxation. Our \na MRF presented here is an extension of the methods applied in our previous work \cite{Gilles2017}. In order to capture the $\text{T}_\text{2,short}^{*}$ dynamics of the \na spin system, several of the works compared in Table \ref{litVals} utilized a 3D radial trajectory \cite{Blunck2018,Lommen2018,Niesporek2017,Nagel2011}, including Kratzer et al. \cite{Kratzer2021} who used a center-out radial trajectory with variable TE, incorporating both single and double echo readouts. In our sequence we utilized the 3D FLORET trajectory \cite{Pipe1999, Pipe2011, Robinson2017, Gilles2017}. FLORET has several advantages over radial acquisition schemes, including improved image quality due to fast gradient spoiling allowing for shorter minimal TR, and improved SNR due to more uniform sampling \cite{Robinson2017}. The goal of our pulse sequence optimization in developing the \na MRF train (Figure \ref{pulseDiagram}) was to best differentiate between GM and WM. This was accomplished by minimizing the PC between the simulated tissue systems using a GA. Our approach resulted in a significantly shorter pulse train than that reported by Kratzer et al. \cite{Kratzer2020, Kratzer2021}, who optimized FAs using the Cramer-Rao lower bound (CRLB) criterion in development of a 1000-pulse train. Our dictionary was generated through direct numerical simulation of the density matrix dynamics using ISTO and incorporated explicit modeling of a $\Delta\text{B}_{1}^{+}$ factor and $\Delta\text{f}_\text{0}$ shift to account for the influence of inhomogeneities directly into the system as described in the Methods section. Kratzer et. al. also used ISTO formalism in their dictionary simulation and included a $\Delta\text{B}_{0}$ correction interpolated along the $\Bar{\Delta\text{B}_{0}}$ axis. These authors extracted relaxation parameters through matching using the highest scalar product between the compressed dictionary and reconstructed images \cite{Kratzer2021}. Our \na MRF matching differs significantly. First, we matched an uncompressed dictionary to reconstructed image data using the PC. Second, we employed a correlation-weighted strategy to include multiple high correlation dictionary matches per voxel.

In our previous work we demonstrated that the evolution of the ISTOs represent the full spin dynamics of the \na spin $\frac{3}{2}$ system, where the simulation of system evolution and relaxation accounts for all single, double and triple quantum coherences (SQC, DQC, TQC). Although multiquantum coherence (MQC) can evolve during a multipulse sequence, and affect the final magnetization \cite{Gilles2017}, the design of our \na MRF pulse sequence did not specifically filter for MQCs. Other methods have been developed for selective filtering of both DQCs \cite{Madelin2014, Tsang2015} and TQCs \cite{Madelin2014}. In each case, careful phase cycling and RF pulse FA selection were designed to select for either DQC or TQC \cite{Madelin2014}. In contrast to this approach, our \na MRF pulse train design was motivated for separating GM and WM signals based on relaxation properties and was developed through combined ISTO simulation of the \na ensemble with GA optimization.

In our \na MRF method, the ISTO simulation that we used to model $\text{T}_\text{2}^{*}$ relaxation included dictionary ranges from 10 to 66 ms for $\text{T}_\text{2,long}^{*}$ and from 0.5 to 66 ms for $\text{T}_\text{2,short}^{*}$. We hypothesized that average values for $\text{T}_\text{2,long}^{*}$ and $\text{T}_\text{2,short}^{*}$ in CSF would be equal or very close to each other. However, the mean $\text{T}_\text{2,short}^{*}$ was significantly shorter than $\text{T}_\text{2,long}^{*}$ in our measurements in CSF. Similar issues with $\text{T}_\text{2}^{*}$ fitting in CSF have been mentioned in the literature \cite{Syeda2019, Blunck2018} where the difficulty arose from using discrete biexponential or monoexponential models. Fingerprint matching, however, operates across a pseudo-continuum and is based on the dynamics of the spin system. Theoretically, this would circumvent some of the shortcomings associated with fitting discrete models. The concentration of higher value  $\text{T}_\text{2,short}^{*}$ in CSF in the center in the central ventricle indicated that partial volume effects from surrounding tissue may have contributed to the low $\text{T}_\text{2,short}^{*}$. Eroding the tissue masks did not significantly change the distribution of $\text{T}_\text{2,short}^{*}$ in CSF, however, any contamination of the ROI could have resulted in a reduced apparent $\text{T}_\text{2,short}^{*}$. 
 
It is also worth noting that our \na MRF method, and the ISTO simulation that we used to generate the fingerprint dictionary, do not make any assumption about either the compartmental homogeneity within the voxel, nor the ratios between relaxation processes for the central and satellite transitions of the \na spins $\frac{\text{3}}{\text{2}}$ (i.e. the long and short relaxation components). Both \na MRF and the ISTO simulation simply estimate average relaxation times in each voxel, which is very likely a volume weighted average of multiple intra-voxel compartments (extracellular and intracellular spaces from multiple cell types) and local magnetic field inhomogeneities. Consequently, there is very little chance to measure the ideal ratio of 0.6:0.4 for $\text{T}_\text{2,short}^{*}$:$\text{T}_\text{2,long}^{*}$ due to quadrupolar relaxation processes. Indeed, if we were to specifically measure the ratio of $\text{T}_\text{2,short}^{*}$:$\text{T}_\text{2,long}^{*}$ with a fitting method (which MRF is not), the ratio of 0.6:0.4 would occur only in a perfectly ideal model where relaxation is purely quadrupolar. This is usually not the case in real samples, including gels, fluids and biological tissues. In reality, dipolar coupling, residual quadrupolar interaction (due to local structural anisotropies), local field inhomogeneities (chemical shift interaction) of various nature, and even spin-orbit interaction where the spin interacts with the magnetic fields generated by the rotational motion of the ion itself, can occur and influence the relaxation process of the \na spins. Such influences effectively act as pertubations to the main quadrupolar relaxation process, and thus generate \na relaxation times that deviate from the ideal case of pure quadrupolar relaxation \cite{Madelin2022, Levitt2008, Kimmich2012}.

\section{Conclusion}

In conclusion, we demonstrated an implementation of \na MRF from a 3D 23-pulse FLORET acquisition that enabled quantitative mapping of $\text{T}_\text{1}$, $\text{T}_\text{2,long}^{*}$, $\text{T}_\text{2,short}^{*}$, SD, $\Delta\text{B}_{1}^{+}$ factor and $\Delta\text{f}_\text{0}$ in about 30 min over the whole brain with 5-mm isotropic resolution at 7 T. Furthermore, we introduced correlation coefficient weighting in the data reconstruction to enhance the robustness of the method and smooth the final maps. 

The proposed \na MRF method could finally be combined with \hh MRF based on our previous work on simultaneous acquisition of \hh MRF and \na MRI \cite{Yu2020,Yu2021,Rodriguez2022} to generate a fully simultaneous \hhna MRF technique \cite{Rodriguez2024}.  

\section*{Acknowledgements}

This work was supported by the NIH/NIBIB grant R01EB026456 and performed under the rubric of the Center for Advanced Imaging Innovation and Research, a NIBIB Biomedical Technology Resource Center (P41EB017183). The authors would like to thank research coordinators Nahbila-Malikha Kumbella and Liz Aguilera, MPH, for volunteer outreach. Some computational requirements for this work were supported by the NYU Langone High Performance Computing (HPC) Core’s resources and personnel.

\section*{Author contributions statement}

L.F.O., G.G.R., Z.Y., G.M. and M.A.C. conceived the experiments. L.F.O and G.G.R. conducted the experiments. L.F.O., G.G.R, G.M and M.A.C. analysed the results. L.F.O, G.M, G.L. and O.D. developed the code for data processing. L.F.O. and G.M. drafted the manuscript. All authors reviewed the manuscript. 

\section*{Financial disclosure}

None.

\section*{Conflict of interest}

G.L. is CEO of MICSI (www.micsi.com), a denoising software based on the Marchenko-Pastur method used in this article. The other authors declare no potential conflict of interests.

\section*{Data availability}

Data analyzed in this study is available on GitHub:  
\url{https://github.com/LaurenFODonnell/Sodium_MRF_Datasets.git}. The Matlab code for sodium NMR dynamics simulation used for generating the fingerprint dictionary is available on Matlab File Exchange: \url{https://www.mathworks.com/matlabcentral/fileexchange/67472-simulation-of-sodium-nmr}. 

\section*{Supplementary material}

Supporting information is available in the Supplementary Material file.

\bibliography{biblio}

\end{document}



\maketitle

\clearpage

\begin{figure*}[h!]
    \centering
    \includegraphics[width=\textwidth]{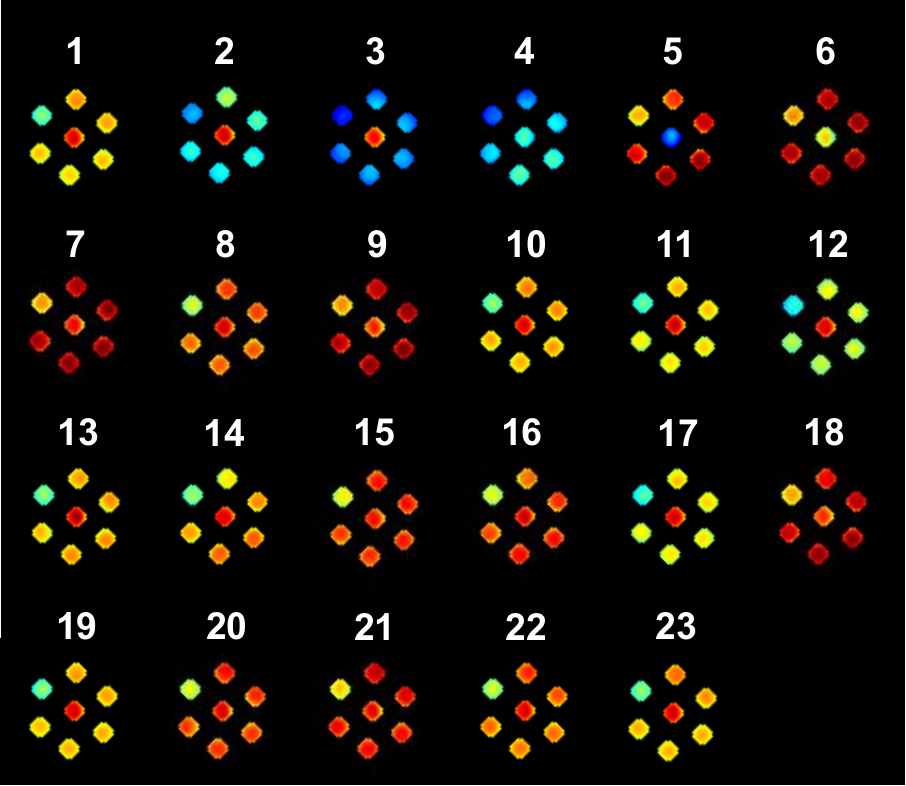}
    \caption{\textbf{Images of the center axial slice of the 7-compartment phantom acquired after each of the 23 pulses in the \na MRF acquisition (global normalization)}. The pulse train is shown in Figure 1(B). These images were normalized to the brightest pixel over all 23 images for the slice shown here. The first three pulses form a 90$^\circ$--180$^\circ$--90$^\circ$ inversion composite block that was used to increase $\text{T}_\text{1}$ sensitivity and improve RF homogeneity of the magnetization inversion. In the composite block, $\tau_{1}$ = $\tau_{2}$ = 7.5 ms. The next 20 pulses form the variable flip angle and phase angle train with $\tau_{3}=...=\tau_{23}$ = 15 ms. Acquisition parameters were TR = 511 ms, isotropic resolution = 5 mm, isotropic FOV = 320 mm, FLORET trajectory (3 hubs/45$^{\circ}$, 100 interleaves/hub) and 16 averages, for a total scan time of 40:52 min.}
    \label{MFlo_phantom}
\end{figure*}

\clearpage

\begin{figure*}[h!]
    \centering
    \includegraphics[width=\textwidth]{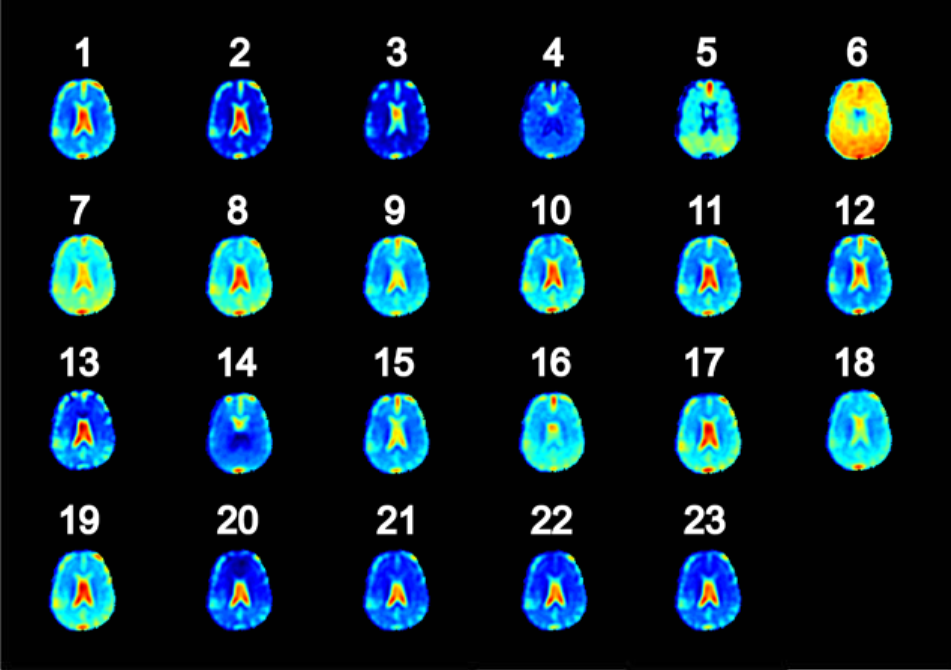}
    \caption{\textbf{Images of the center axial slice of the brain of volunteer 5 acquired after each of the 23 pulses in the \na MRF acquisition (individual normalization).} The pulse train is shown in Figure 1(B). These images were normalized to the brightest pixel in each of the 23 individual images for the slice shown here. The first three pulses form a 90$^\circ$--180$^\circ$--90$^\circ$ inversion composite block that was used to increase $\text{T}_\text{1}$ sensitivity and improve RF homogeneity of the magnetization inversion. In the composite block, $\tau_{1}$ = $\tau_{2}$ = 7.5 ms. The next 20 pulses form the variable flip angle and phase angle train with $\tau_{3}=...=\tau_{23}$ = 15 ms. Acquisition parameters were TR = 702 ms, isotropic resolution = 5 mm, isotropic FOV = 320 mm, FLORET trajectory (3 hubs/$45^{\circ}$, 100 interleaves/hub), 4 scans, 2 averages/scan, for a total acquisition time of 31:06 min.}
    \label{MFlo_Brain}
\end{figure*}

\clearpage

\begin{figure*}[h!]
    \centering
    \includegraphics[width=\textwidth]{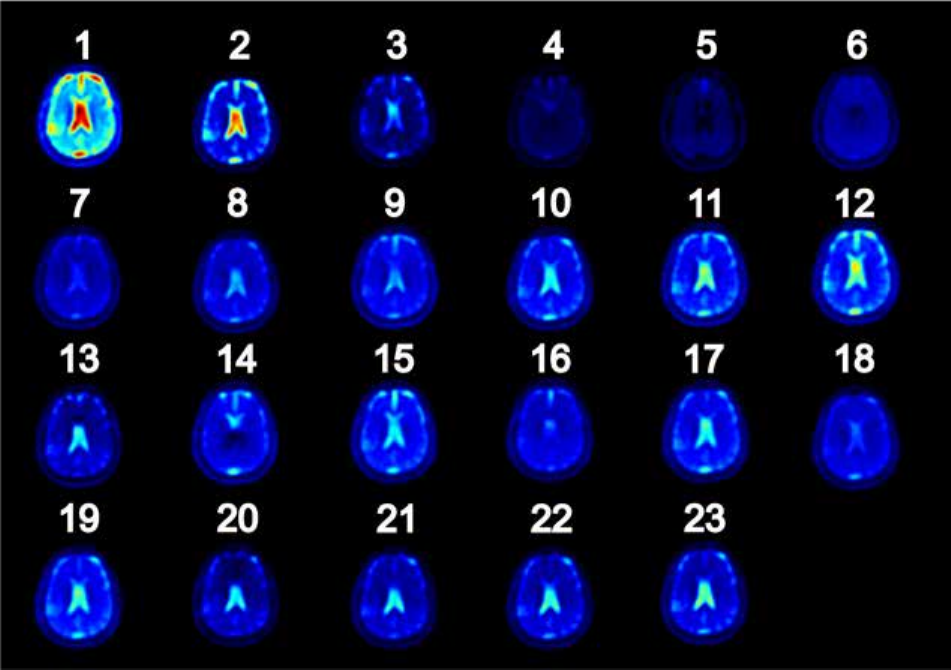}
    \caption{\textbf{Images of the center axial slice of the brain of volunteer 5 acquired after each of the 23 pulses in the \na MRF acquisition (global normalization).} The pulse train is shown in Figure 1(B). These images were normalized to the brightest pixel over all 23 images for the slice shown here. The first three pulses form a 90$^\circ$--180$^\circ$--90$^\circ$ inversion composite block that was used to increase $\text{T}_\text{1}$ sensitivity and improve RF homogeneity of the magnetization inversion. In the composite block, $\tau_{1}$ = $\tau_{2}$ = 7.5 ms. The next 20 pulses form the variable flip angle and phase angle train with $\tau_{3}=...=\tau_{23}$ = 15 ms. Acquisition parameters were TR = 702 ms, isotropic resolution = 5 mm, isotropic FOV = 320 mm, FLORET trajectory (3 hubs/45$^{\circ}$, 100 interleaves/hub), 4 scans, 2 averages/scan, for a total acquisition time of 31:06 min.}
    \label{MFlo_Brain2}
\end{figure*}

\clearpage

\begin{figure*}[h!]
    \centering
    \includegraphics[width=\textwidth]{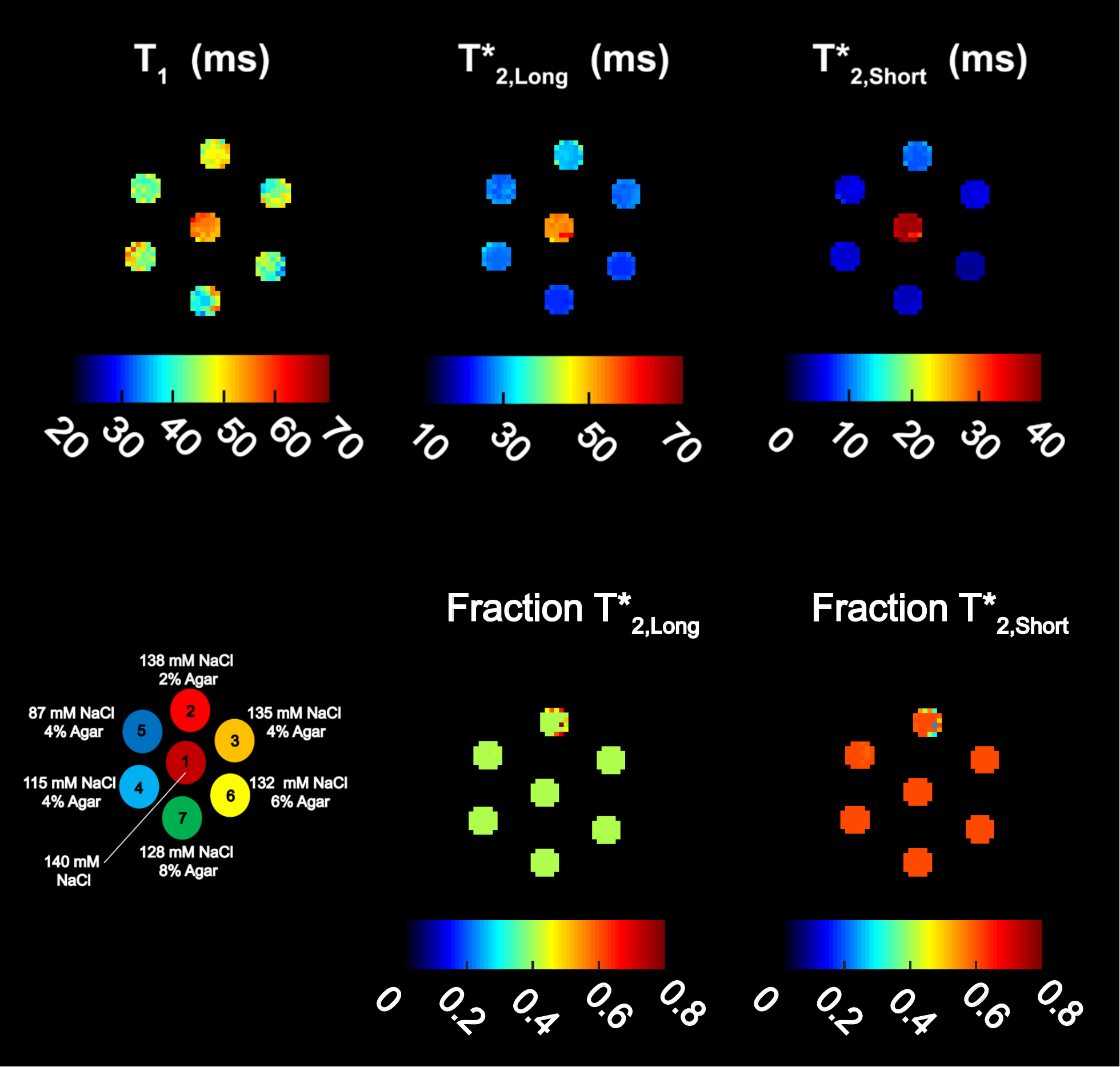}
    \caption{\textbf{Reference method (RM) maps of the 7-compartment phantom.} Phantom images from FLORET for $\text{T}_\text{1}$ and $\text{T}_\text{2}^*$ reference experiments were reconstructed in the same way as the \na MRF data. Curve fitting was applied voxelwise over the axial plane in the central slice of the phantom using the Levenberg-Marquardt algorithm applied using \textit{lsqcurvefit} in MATLAB. See Equations 7 and 8 in the main text for the fitting kernels. The signal fractions for $\text{T}_\text{2,short}^{*}$ ($f$) and $\text{T}_\text{2,long}^{*}$  ($1-f$) are also presented. The mean values for $\text{T}_\text{1}$, $\text{T}_\text{2,long}^{*}$, $\text{T}_\text{2,short}^{*}$ in individual ROIs are given in Table 3.}
    \label{MFlo_Brain2}
\end{figure*}

\clearpage

\begin{figure*}[h!]
    \centering
    \includegraphics[width=\textwidth]{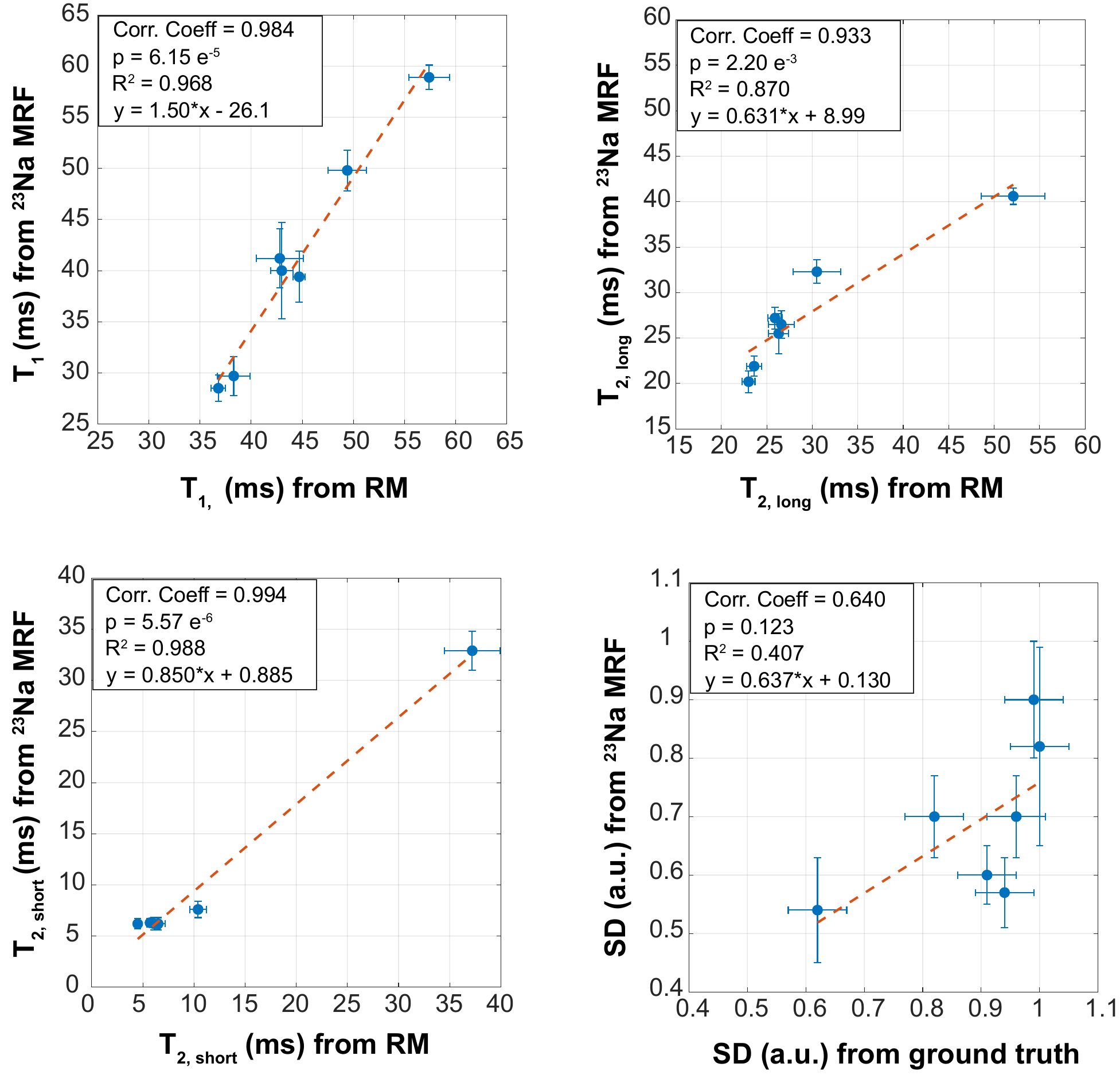}
    \caption{\textbf{Scatter plots of $\text{T}_\text{1}$, $\text{T}_\text{2,long}^{*}$, $\text{T}_\text{2,short}^{*}$ and SD measured with \na MRF versus the reference method (RM).} Mean values and standard deviation (STDV) for $\text{T}_\text{1}$, $\text{T}_\text{2,long}^{*}$, $\text{T}_\text{2,short}^{*}$ are given in Table 1. The SD reference values were taken from the ground truth in the 7-compartment phantom composition with $\pm$5\% STDV. The error bars correspond to $\pm$1 STDV.}
    \label{Scatter}
\end{figure*}

\clearpage

\begin{figure*}[h!]
    \centering
    \includegraphics[width=\textwidth]{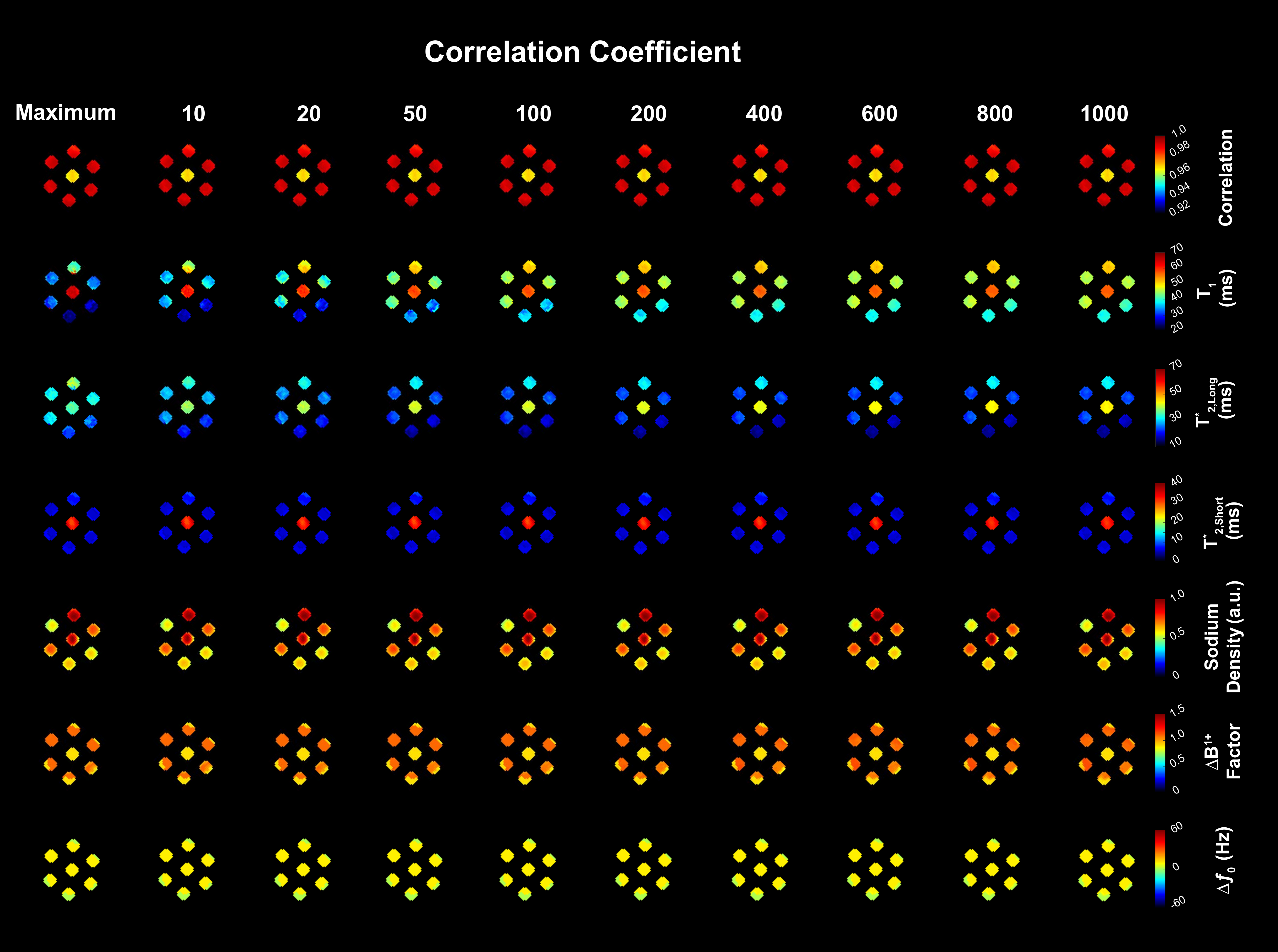}
    \caption{\textbf{\na MRF maps produced after matching with different numbers of correlation coefficients (k) used as average weighting factors in a 7-compartment phantom.} The "Maximum" label above the left column indicates the result of mapping between the fingerprint dictionary and the image signal using only the highest correlation coefficient (k = 1). The values above each column indicate the number of highest correlation coefficients included in the map calculations. The maps for k = \{10, 20, ..., 1000\} therefore correspond to the weighted average of signals with the k highest correlation coefficients, where the value of the correlation coefficient itself was used as the weighting factor.}
    \label{CorrWeight_phantom}
\end{figure*}

\clearpage

\begin{figure*}[h!]
    \centering
    \includegraphics[width=0.90\textwidth]{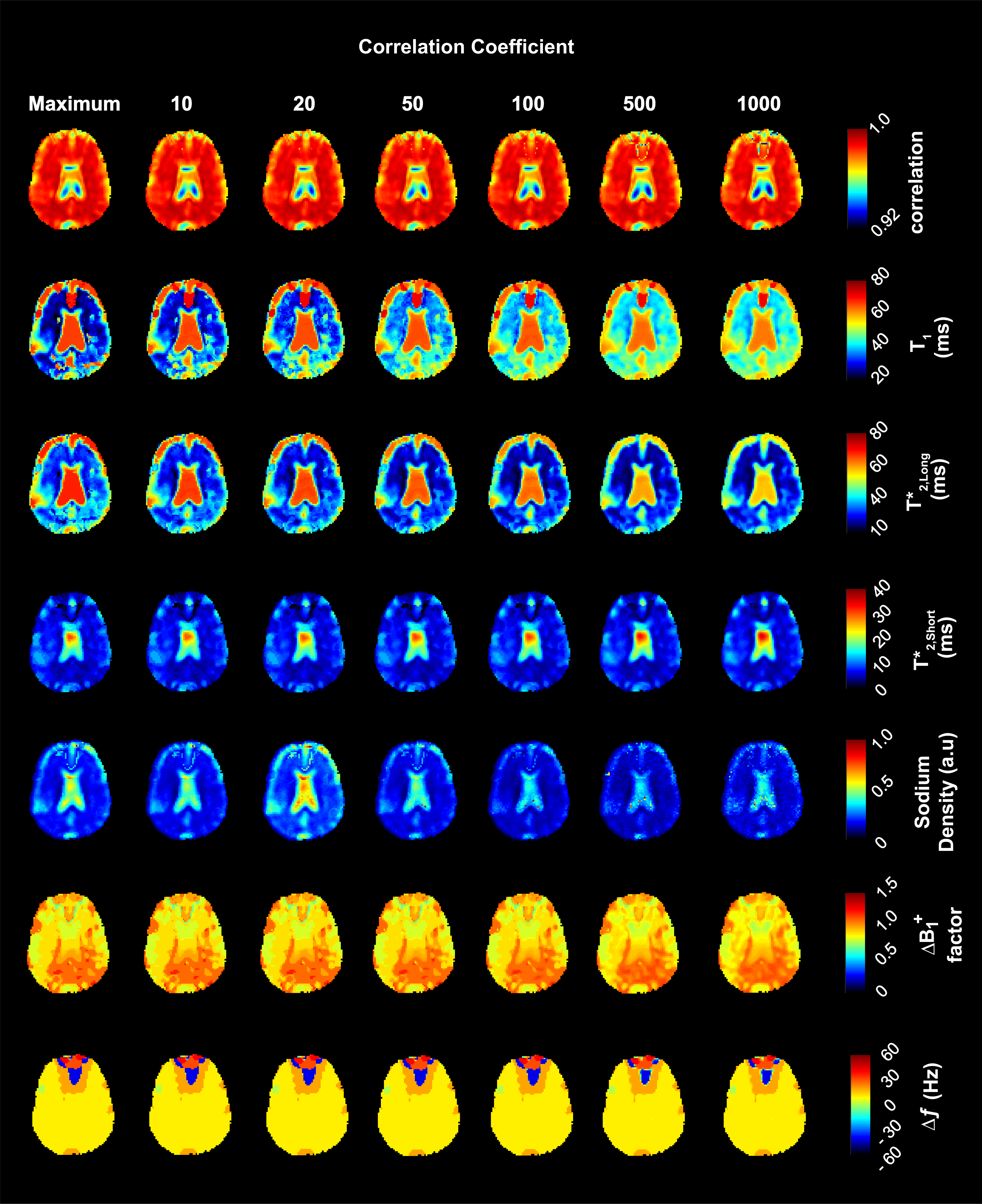}
    \caption{\textbf{\na MRF maps produced after matching with different numbers of correlation coefficients (k) used as weighting factors in brain for volunteer 5.}  The "Maximum" label above the left column indicates the result of mapping between the fingerprint dictionary and the image signal using only the highest correlation coefficient (k = 1). The values above each column indicate the number of highest correlation coefficients included in the map calculations. The maps for k = \{10, 20, ..., 1000\} therefore correspond to the weighted average of signals with the k highest correlation coefficients, where the value of the correlation coefficient itself was used as the weighting factor.}
    \label{CorrWeight_Brain}
\end{figure*}

\clearpage

\begin{figure*}[h!]
    \centering
    \includegraphics[width=\textwidth]{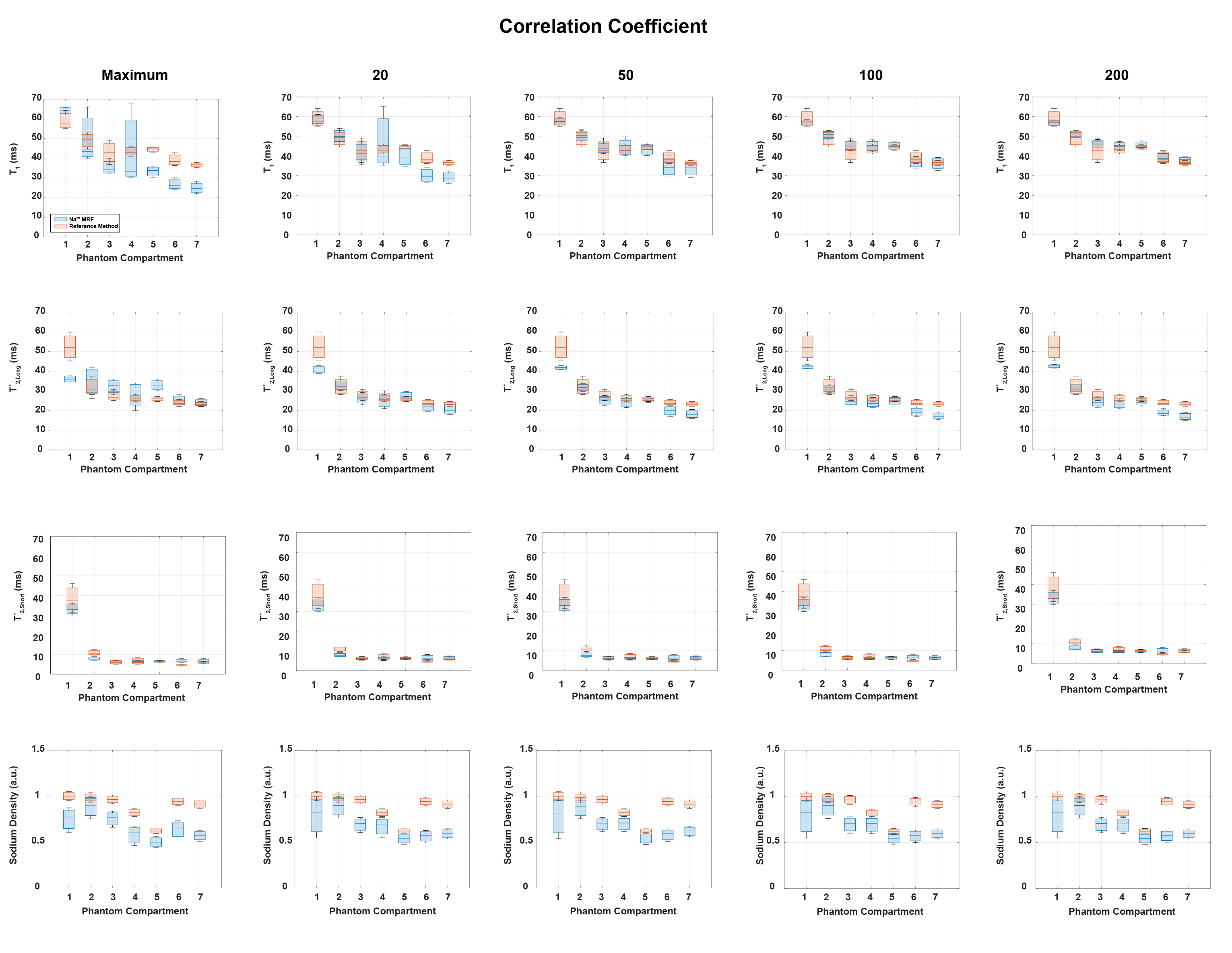}
    \caption{\textbf{Boxplots of $\text{T}_\text{1}$, $\text{T}_\text{2,long}^{*}$, $\text{T}_\text{2,short}^{*}$ and SD measured in the 7-compartment phantom.} Blue boxes represent data from \na MRF. Red boxes represent data from the reference method (RM) for relaxation times, and from ground truth for SD. Boxplots are grouped by column for k = \{1 (maximum), 20, 50, 100, 200\} highest correlation coefficients used for the weighed average of each map parameter.}
    \label{BoxPlotsAll_phantom}
\end{figure*}

\clearpage

\begin{figure*}[h!]
    \centering
    \includegraphics[width=0.75\textwidth]{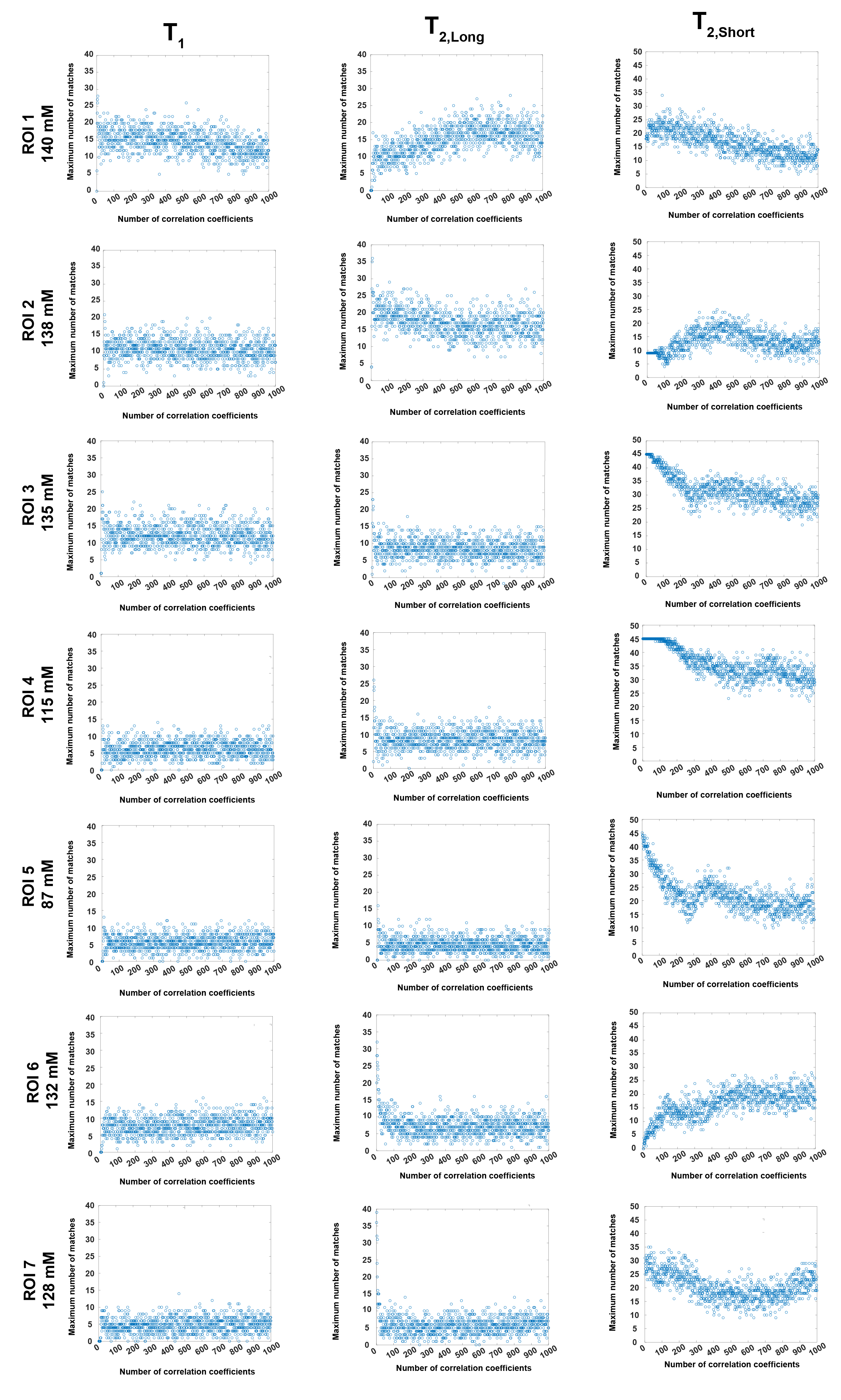}
    \caption{\textbf{Maximum number of fingerprint dictionary matches in the 7-compartment phantom.} The plots show the maximum number of fingerprint dictionary matches made to the subset of pixels within a reference range for $\text{T}_\text{1}$, $\text{T}_\text{2,long}^{*}$ and $\text{T}_\text{2,short}^{*}$ versus the number of k = \{1, 2, 3 ..., 1000\} correlation coefficients included in the matching. The limits were imposed using the reference method (values are shown Table 1).}
    \label{MaxCorrGraph_Phan}
\end{figure*}

\clearpage

\begin{figure*}[h!]
    \centering
    \includegraphics[width=\textwidth]{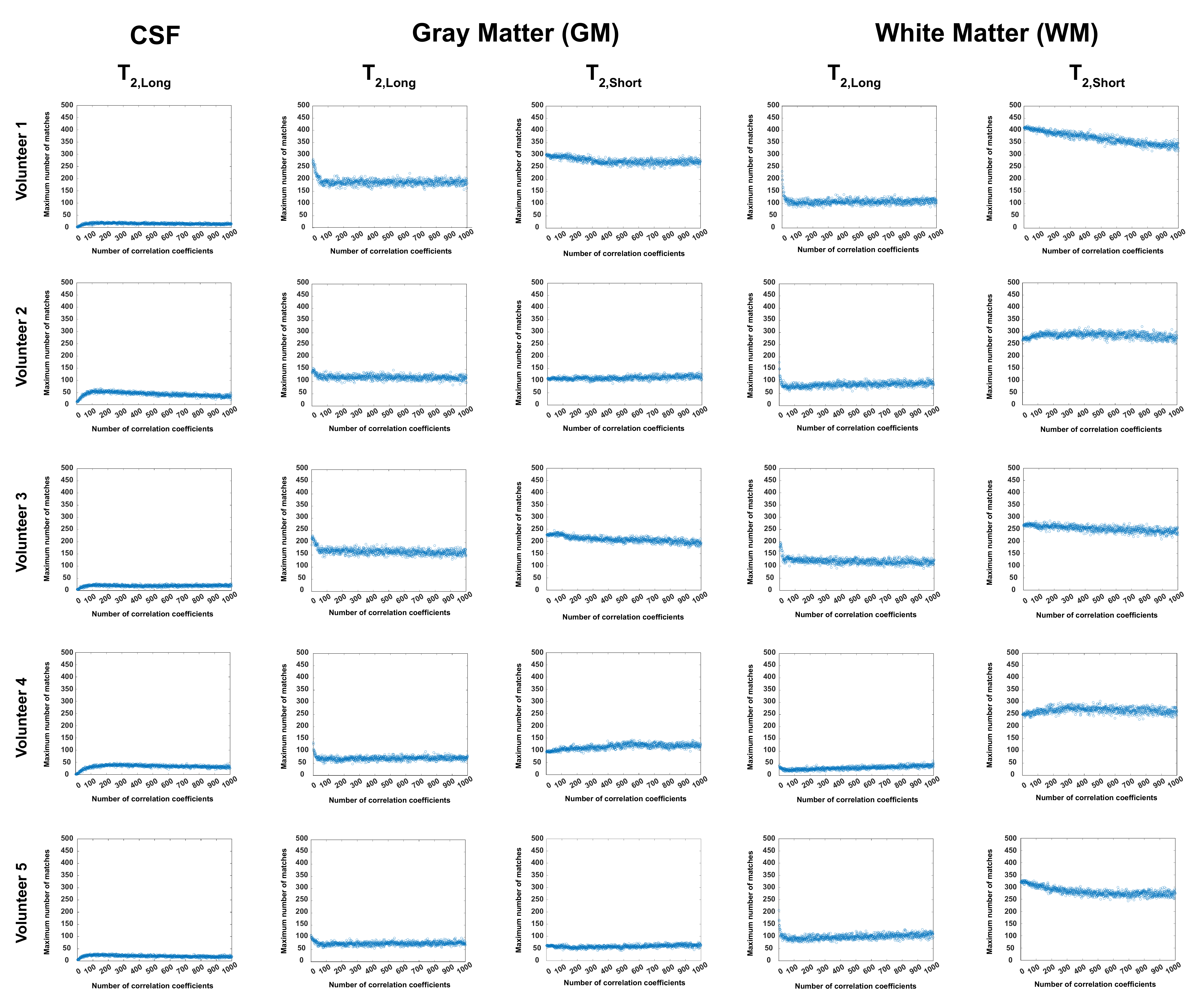}
    \caption{\textbf{Maximum number of fingerprint dictionary matches in brain.} The plots show the maximum number of fingerprint dictionary matches made to the subset of pixels within a reference range for $\text{T}_\text{2,long}^{*}$ in CSF, GM and WM and $\text{T}_\text{2,short}^{*}$ in GM and WM versus the number of k = \{1, 2, 3 ..., 1000\} correlation coefficients included in matching. Ranges for relaxation were determined by literature values shown in Table 3. We omitted $\text{T}_\text{1}$ from this analysis because we did not find enough reports of \na $\text{T}_\text{1}$ at 7 T in the literature to impose limits for our computation.}
    \label{boxPlPhan}
\end{figure*}

\clearpage

\begin{figure*}[h!]
    \centering
    \includegraphics[width=\textwidth]{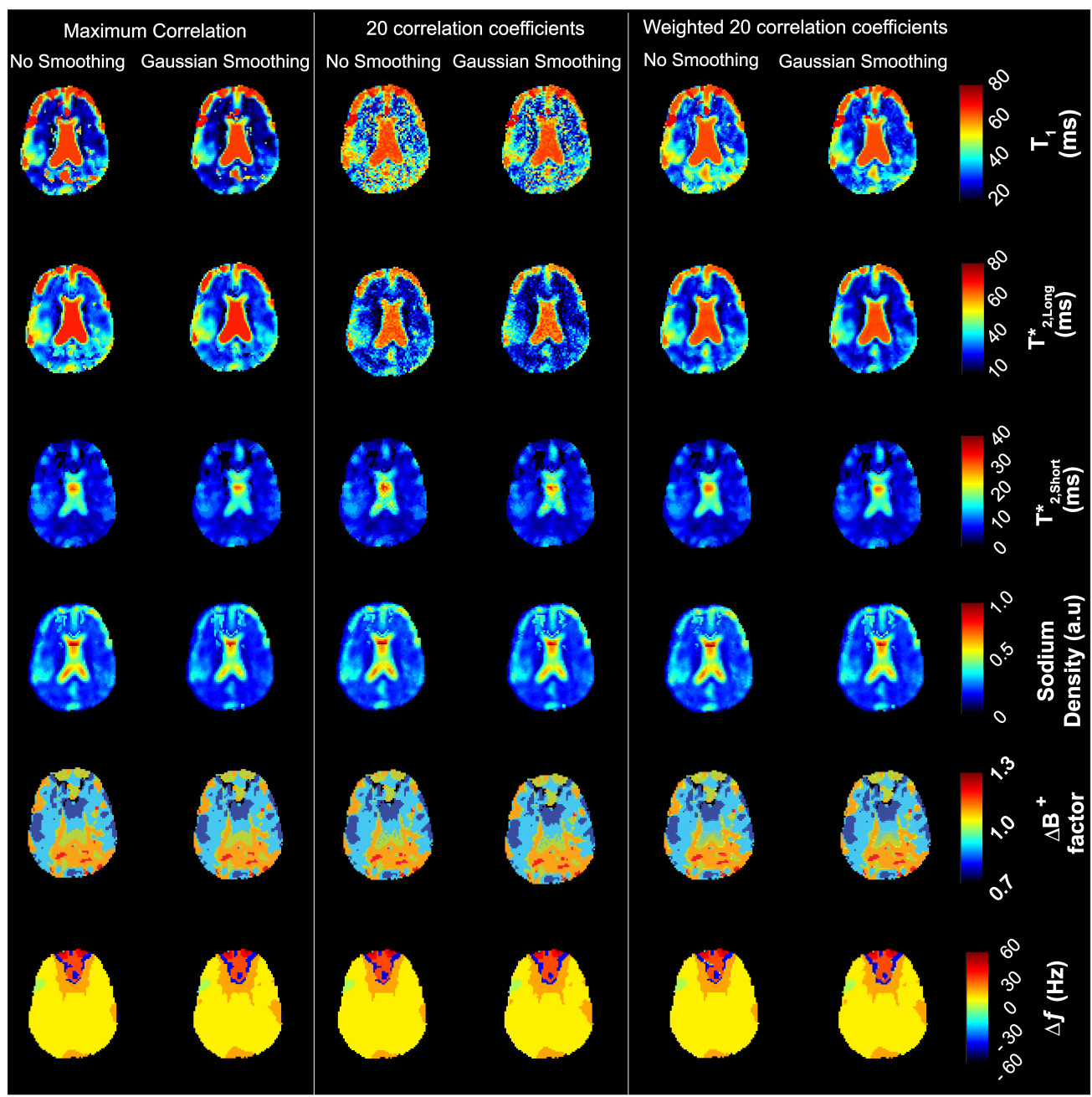}
    \caption{\textbf{Comparison of brain maps for a single axial slice reconstructed with and without Gaussian filtering (volunteer 5).} Gaussian filtering (smoothing) was applied prior to matching using a heuristically determined window executed in MATLAB. The filter was applied over individual slices, in plane, on the image acquired after each of the 23 pulses of the \na MRF pulse train. After filtering, fingerprint dictionary matching was carried out as described in the Methods section. The maps shown here correspond to: signal with maximum correlation coefficient (k = 1) only, with and without filtering; average over k = 20 signals with the highest correlation coefficients, but without the correlation coefficient weighting (simple averaging with weighting = 1 for all k signals), with and without filtering; and average over k = 20 signals with the highest correlation coefficients, with correlation coefficient weighting, with and without filtering.}
    \label{gSmooth_Brain}
\end{figure*}
